\documentclass[twocolumn]{aa}
\usepackage{graphicx}
\usepackage{longtable}
\usepackage{supertabular}

\usepackage{natbib}
\bibpunct{(}{)}{;}{a}{}{,}
\def\vsini{$V\!\sin i$}
\def\teff{$T_{\rm eff}$}
\def\tpole{$T_{\rm polar}$}
\def\rpole{$R_{\rm p}$}

\def\logg{$\log~g$}
\def\om{$\Omega$}
\def\omc{$\Omega/\Omega_{\rm{c}}$}
\def\vevc{${\rm V_e}/{\rm V_c}$}
\def\kps{km~s$^{-1}$}
\def\dvsini{$\delta\!V\!\sin i$}
\def\top{$T_{\rm eff}^{\rm o}$}
\def\gop{$\log g_{\rm o}$}
\def\vsinio{$V\!\sin i$}
\def\vsiniap{$V\!\sin i_{\rm~\!app.}$}
\def\tap{$T_{\rm eff}^{\rm app.}$}
\def\gap{$\log g_{\rm app.}$}

\begin{document}
\title{Effects of gravitational darkening on the determination of fundamental
parameters in fast rotating B-type stars
}


\author{
Y. Fr\'emat \inst{1} \and J. Zorec \inst{2} \and A.-M. Hubert
\inst{3} \and M. Floquet \inst{3} } \offprints{yves.fremat@oma.be}
\institute{Royal Observatory of Belgium, 3 avenue circulaire,
B-1180
 Bruxelles\\
              \email{yves.fremat@oma.be}
\and
Institut d'Astrophysique de Paris, CNRS, 98bis Boulevard Arago,
F-75014 Paris
\and
Observatoire de Paris, Section d'Astrophysique de Meudon, GEPI,
FRE K 2459, 5 Place Jules Janssen, F-92195 Meudon CEDEX
}
\date{--}
\abstract{In this paper we develop a calculation code to account
for the effects carried by fast rotation on the observed spectra
of early-type stars. Stars are assumed to be in rigid rotation and
the grid of plane-parallel model atmospheres used to represent
the gravitational darkening are calculated by means of a non-LTE
approach. Attention is paid on the relation between the {\it
apparent} and {\it parent non-rotating counterpart} stellar
fundamental parameters and apparent and true \vsini\ parameters
as a function of the rotation rate $\Omega/\Omega_{\rm c}$,
stellar mass and inclination angle. It is shown that omission of
gravitational darkening in the analysis of chemical abundances of
CNO elements can produce systematic overestimation or
underestimation, depending on the lines used, rotational rate and
inclination angle. The proximity of Be stars to the critical
rotation is re-discussed by correcting not only the \vsini\ of
130 Be stars, but also { their effective temperature and gravity
to account for stellar rotationally induced geometrical
distortion and for the concomitant gravitational darkening
effect}. We concluded that the increase of the \vsini\ estimate is
accompanied by an even higher value of the stellar equatorial
critical velocity, so that the most probable average rate of
angular velocity of Be stars attains $\Omega/\Omega_{\rm c}
\simeq$ 0.88.\par

 }
\authorrunning{Y. Fr\'emat et al.}
\titlerunning{Effects of fast rotation}

\maketitle

\section{Introduction}

Fast rotation produces a polar flattening and an equatorial
stretching of stars, which in turn induce non-uniform surface
gravity and temperature distributions: the gravity darkening
effect \citep[GD, ][]{1924MNRAS..84..665Z}. Owing to the
levitation effect introduced by the rotation, rotating stars
evolve as they had a lower mass: their core density is higher,
the total bolometric luminosity produced is lower and they spend
a longer time on the main sequence evolutionary phase than slowly
rotating stars with the same mass \citep{1970A&A.....8...76S,
1971ApJ...167..153B,1979ApJ...230..230C,1982RPPh...44..831M,
2000A&A...361..101M}. The geometrical deformation of the star and
the GD further influence the apparent status of the star, which
also depends on whether it has a radiative or convective envelope
\citep{1967ZA.....65...89L,1972ApL....10..175T,1974MNRAS.167..199C,
1998A&A...335..647C,2000A&A...359..289C}. As predicted by models
of stellar evolution
\citep{1979ApJ...232..531E,1991ApJ...367..239P,
2000ApJ...544.1016H,2000A&A...361..101M}, fast rotation is further
expected to generate several instabilities that contribute to
redistribute the internal angular momentum. They induce thus a
turbulent diffusion in the stellar interior
\citep{1992A&A...265..115Z}, which in massive objects drives the
CNO-cycled material from the core to the envelope and changes the
atmospheric chemical composition in these stars
\citep{2000A&A...361..101M}.\par

\subsection{Rotation dependent spectra of early type stars}

 Radiative and hydrostatic equilibrium conditions in atmospheres of rotating
early-type stars impose in each point of the stellar surface the
emitted bolometric flux be proportional to the local gravity $F
\propto g_{\rm eff}^{\beta}$ with $\beta\simeq$ 1 if $T_{\rm eff}
\ga$ 7000 K and $g_{\rm eff} = g_{grav}-g_{rot}+g_{rad}$
\citep{1924MNRAS..84..665Z, 2000A&A...359..289C}. Although a
non-local theory of radiative transfer in rotating stars does not
reproduce von Zeipel's relation, meridional circulation and
turbulence in surface layers recover the validity of this law
\citep{1974MNRAS.167..199C}. For B-type star effective
temperatures, the radiation pressure $g_{rad}$ can be neglected.
Energy distributions, photometric indices and spectral lines
produced in gravity-darkened early type rotating stars were
calculated by several authors, either for rigid
\citep{1963ApJ...138.1134C,1965ApJ...142..265C,1966ApJ...146..914C,
1968ApJ...151..217C,1968ApJ...152..847C,1974ApJ...191..157C,
1968ApJ...151.1057H,1972A&A....17..161H,1968ApJ...153..465H,
1971A&A....13..353H,1977ApJS...34...41C,1991ApJS...77..541C,
1970A&A.....7..120M,1972A&A....21..279M} or for differential
rotators with conservative internal rotational laws
\citep{1985MNRAS.213..519C, 1986serd.book.....Z}. A0 to F5-type
rigid rotators were studied by \citet{1999A&A...346..586P}. The
effects of the GD on the $V\!\sin i$ determinations were explored
by \citet{1968MNRAS.140..149S} for rigid rotators and Zorec et
al. (1988) for differential rotators. \citet{1968MNRAS.140..149S}
concluded that the reduced contribution to the He\,{\sc i} 4471
line broadening in the equatorial region may lead to
underestimated rotational parameters in early-type stars, while
Zorec et al. (1988a) obtained double valued $V\!\sin i$ in star
models with polar hollows. Recently, \citet{2004MNRAS.350..189T}
rediscussed the GD effect on the $V\!\sin i$ values of rigid
early-type rotators and concluded that classic $V\!\sin i$
determinations for B0 to B9-type stars can be underestimated by
12 to 33\% at $\omega = \Omega/\Omega_c =$ 0.95 rotational rates,
if the He\,{\sc i} 4471 line is used and from 9 to 17\%, if the
Mg\,{\sc ii} 4481 line is studied.\par

\subsection{Aim of the present work}

 Apart from the obvious rotation dependent Doppler effect, four factors, at
least, still concur to produce the observed line broadening in
gravity darkened fast rotators: a) changes with the surface
non-uniform temperature of the continuous spectrum specific
intensity; b) dependence of the intrinsic lines equivalent width
with the local temperature and gravity; c) stretching of the
isotemperature and isogravity regions, which make that in strips
of constant radial velocity for rather large Doppler
displacements there can be temperatures close to that of the
undistorted star; d) for high enough inclinations, the flattened
stellar disc produce shortened constant Doppler displacement
strips that tend to shallow the rotationally broadened line
profile. A look on the He\,{\sc i} 4471 and Mg\,{\sc ii} 4481
line equivalent width dependence with temperature and gravity can
be obtained in \citet{1982A&AS...50..199D} and Sect.
\ref{sec:prv}. These lines are currently used to determine the
$V\!\sin i$ parameters of early-type stars
\citep{1975ApJS...29..137S}. From classic model atmospheres we
obtain that at the He\,{\sc i} 4471 and Mg\,{\sc ii} 4481 line
wavelengths the continuum specific intensity scales as $I_{\rm
c}\sim$ $T_{\rm eff}^a$, with $a \sim 2.2$ if $T_{\rm eff} \ga$
7500 K. Since is it $T_{\rm eff}({\rm eq})/T_{\rm eff}({\rm
pole}) =$ $(g_{\rm eq}/g_{\rm pole})^{0.25}$, we infer that for
$\omega = 0.95$ the equatorial regions contribute less to the
flux of the broadened line profile by a factor $I_{\rm c}({\rm
eq})/I_{\rm c}({\rm pole})\la$ 0.2 than in the pole. Line
profiles are then expected to be narrower than those calculated
with models where GD is neglected. At high aspect angles, effect
d) compensates partially this deepening. Depending on the
temperature of the star, in fast rotators the change of the local
$T_{\rm eff}$ and $\log g$ with latitude produce the equivalent
width of the He\,{\sc i} 4471 line be either $W_{He {\sc i}}({\rm
eq})
> W_{He {\sc i}}({\rm pole})$, or $W_{He {\sc i}}({\rm eq}) < W_{He {\sc i}}({\rm pole})$,
while the equivalent width of the Mg\,{\sc ii} 4481 line is always
$W_{Mg {\sc ii}}({\rm eq}) > W_{Mg {\sc ii}}({\rm pole})$. Let us
note that for the \ion{He}{i} 4471 line, we actually mean the
blend \ion{He}{i} 4471+\ion{He}{i} 4470, where the second line is
a forbidden component and has a somewhat different sensitivity to
fundamental parameters than the permitted $\lambda4471$
component. For the He\,{\sc i} 4471 and Mg\,{\sc ii} 4481 lines
also respond in a different way to the GD and because they
overlap partially, the $V\!\sin i$ determination demands that we
proceed to fit both lines simultaneously.\par
 The aim of this paper is thus to take into account, as properly as
possible, the changes of the stellar radii $R_{\rm pole}$ and
$R_{\rm eq}$ and of the global bolometric luminosity with
$\omega$ in the calculation of the emitted radiation fluxes. We
will also fit both He\,{\sc i} 4471 and Mg\,{\sc ii} 4481 lines
in order to determine up to what degree the \vsini\ parameters of
early-type fast rotators are underestimated if they are obtained
assuming that rotational distortion and GD can be neglected. This
is of particular interest for Be stars to know whether they are
near-critical rotators \citep{2004MNRAS.350..189T}. In this
paper, we will also discuss the effect of the rapid rigid
rotation on the determination of the stellar effective
temperature, surface gravity and chemical composition.\par

\section{Model atmospheres}
\label{sec:ma}

 Fast rotation flattens the star and produces non-uniform density
and temperature distributions in its surface. To take into
account the first order effects of this flattening on the stellar
spectrum of B-type stars, we adopted a method similar to the one
described by \citet{1991ApJS...77..541C}, but modified to
introduce the changes detailed in Sect. \ref{sec:flat}. In this
approach, that we applied in a computer code hereafter named {\sc
fastrot}, the stellar photosphere is replaced by a mesh of
plane-parallel model atmospheres depending each on the local
temperature and surface gravity (Sect. \ref{sec:plan}).\par

\subsection{Flattening and gravitational darkening}
\label{sec:flat}

 We assume that the stars are rigid rotators without any surface latitudinal
differential rotation component. We adopted the Roche
approximation for the stellar surface equipotentials, i.e. in the
gravity potential the dipole and higher order terms are neglected
\citep{1978trs..book.....T}. In the expression of the surface
effective gravity we also neglected the latitude-dependent
radiation pressure term
\citep[$g_{rad}=0$;][]{2000A&A...361..159M}. From the Roche
equipotentials it follows that the ratio between the equatorial
to the polar radii at the rotational rate $\omega =
\Omega/\Omega_{\rm c}$ ($\Omega_{\rm c}$ = critical angular
velocity) is given by:

\begin{equation}
\left.\begin{array}{lcl}
\frac{R_{\rm e}(\omega)}{R_{\rm p}(\omega)} & = & 1+\frac{1}{2}\eta \\
\eta & = & \omega^2(\frac{R_{\rm e}(\omega)}{R_{\rm c}})^3
\end{array}
\right\} \label{eq1}
\end{equation}

\noindent where $R_{\rm c}$ is the equatorial radius at critical
rotation $(\omega = 1)$.

{ In what follows we use the \omc\ to characterize the rotation
rate instead of \vevc\, where V$_{\rm e}$ and V$_{\rm c}$ are the
actual and critical linear equatorial rotation velocities
respectively. The main reason for this choice is that as we
assume stars are rigid rotators, the angular velocity $\Omega$,
which is the same over the whole star (surface and interior) and
independent of any other stellar quantity, it can be considered
as an independent fundamental parameter. We note, however, that
the stellar rotational distortion and the associated gravity
darkening effect are controlled by the parameter $\eta$ given in
(\ref{eq1}). Though \omc, \vevc\ and $\eta$ are functions having
the same germs 0 and 1, it is \omc\ $<$ \vevc\ $<~\eta$ in the
open interval $0 <$ \omc\ $< 1$.}\par

{ In previous works it is often assumed that} $R_{\rm p}(\omega)
=$ $R_{\rm o}$ where $R_{\rm o}$ is the radius of the star at
rest \citep{1991ApJS...77..541C,A&A...408..707R,
2004MNRAS.350..189T}. In the present study we adopted instead the
following interpolation expression for $R_{\rm p}$, derived from
the calculations performed by \citet{1971ApJ...167..153B},
\citet{1979ApJ...230..230C} and \citet{1988CRASM.306.1265Z}:

\begin{equation}
\left.\begin{array}{lcl}
\frac{R_{\rm p}(\omega)}{R_{\rm o}} & = & 1-P(M)\tau \\
P(M) & = & 5.66+\frac{9.43}{M^2}, \ \ \ \ \ {\rm for}\ M \ga 2 \\
\end{array}
\right\} \label{eq2}
\end{equation}

\noindent where $M$ is the stellar mass in solar units and $\tau
= K/|W|$ with $K$ the rotational kinetic energy and $W$ the
stellar gravitational potential energy. A rough relation between
$\tau$ and $\eta$ is given by $\tau \simeq$
$[0.0072+0.008\eta^{1/2}]\eta^{1/2}$. { Although (\ref{eq1})
implies $R_{\rm c} =$ 1.5$R_{\rm o}$ at critical rotation, as
from (\ref{eq2}) $R_{\rm p}(\omega) <$ $R_{\rm o}$, depending on
the B-type stellar mass, it is $R_{\rm c} =$ 1.3$R_{\rm o}$ to
1.4$R_{\rm o}$, rather than $R_{\rm c} =$ 1.5$R_{\rm o}$ as
usually assumed in the literature.}\par

 The total bolometric luminosity produced in the stellar core was assumed
given by:

\begin{equation}
L(\omega,M,g_{\rm o}) = L_{\rm o}(M,g_{\rm o})f_{\rm L}(\tau,M)
\label{eq3}
\end{equation}

\noindent where $L_{\rm o}$ is the bolometric luminosity of an
homologous non-rotating star with mass $M$ and rest surface gravity
$g_{\rm o}$. The expression for the factor $f_{\rm L}$ was
derived from Clement's (1979) models for $M \ga 1.5$:

\begin{equation}
\left.\begin{array}{lcl}
f_{\rm L}& = & a(M)+[1-a(M)]e^{-b(M)\tau}\\
a(M) & = & 0.675+0.046M^{1/2}  \\
b(M) & = & 52.71+20.63M^{1/2}  \\
\end{array}
\right\} \label{eq4}
\end{equation}

 The quoted models (Bodenheimer 1971, Clement 1979, Zorec et al. 1988b) were
calculated for stars with internal conservative differential
rotation laws and for energy ratios $\tau$ that reach the global
secular stability limit $\tau = K/|W| \simeq  0.14$ and even
higher. In all these models the core is supposed to rotate as a
rigid body. Moreover, these calculations show that the radii
ratio $R_{\rm p}(\omega)/R_{\rm c}$ and the function $f_{\rm L}$
do not depend on the rotation law in the stellar envelope
\citep{1986serd.book.....Z}. Nevertheless, the interpolation
relations (\ref{eq2}) and (\ref{eq4}) are for $0 \leq \tau=K/|W|
\la 0.03$, range of $\tau$ values which are suited for rigid
rotators, including the rigid critical rotation, where $\tau =
K/|W| \la 0.02$ whatever the stellar mass in the $2 \la M \la 60$
range.\par
 Relation (\ref{eq3}) takes into account only the change of the
total bolometric luminosity due to the mechanic rotationally
induced changes of the stellar core temperature and pressure
\citep{1970A&A.....8...76S}. { Whenever possible, we have checked
that the changes introduced by mixing phenomena induced by the
instabilities produced by the rotation \citep{1976ApJ...210..184E,
1996A&A...313..140M,2000A&A...361..101M} do not introduce
significant changes to the $f_{\rm L}$ given by \ref{eq4}.}\par

 In this work the GD law was written as:

\begin{equation}
T_{\rm eff}^4(\theta,\omega) = \gamma(\omega,M,R_{\rm
o})g^{\beta}(\theta,\omega) \label{eq5}
\end{equation}

\noindent where $\theta$ is the colatitude angle; $M$ is the stellar mass;
$R_{\rm o}$ is the radius of the star at rest; $g$ is the modulus of the
gravity vector $\overline{g}=-\overline{\nabla}\Phi$ with $\Phi$ being the
Roche potential; $\beta$ was considered $= 1$ for all local $T_{\rm
eff}(\theta,\omega)\ga$ 7000 K. For lower local \teff, $\beta$ was
interpolated in Claret (1998). The factor $\gamma$ was calculated for each
$\omega$ as follows:

\begin{equation}
\gamma = \frac{L(\omega,M,g_{\rm o})}{\sigma\int_{\cal
S}g^\beta{\rm d}{\cal S}}. \label{eq6}
\end{equation}

\noindent where $\sigma$ is the \v Stefan-Boltzmann constant and
${\cal S}$ is the total area of the surface Roche equipotential.
From (\ref{eq3}) it comes that the global effective temperature
$\overline{T_{\rm eff}} =$ $[L(\omega )/\sigma{\cal S}]^{1/4}$ is
smaller than $T_{\rm eff}^{\rm o} =$ $[L_{\rm o}/4\pi\sigma
R_{\rm o}^2]^{1/4}$, the effective temperature of a non-rotating
star with the same mass. On the other hand, relations (\ref{eq5})
and (\ref{eq6}) imply that $T_{\rm eff}(\omega)_{\rm pole}$ $>$
$\overline{T_{\rm eff}}$. In the literature relation (\ref{eq5})
is currently replaced by $T_{\rm eff}^4(\theta,\omega) =$ $T_{\rm
eff}^4(\theta\!\!=\!\!0,\omega
)[g(\theta,\omega)/g(\theta\!\!=\!\!0,\omega)]^{\beta}$, which
hinders to relate the apparent stellar fundamental parameters
with { those} the star would had if it were at rest. We note that
the relations $T_{\rm eff}(\theta\!\!=\!\!0,\omega) =$ $T_{\rm
eff}^{\rm o}$ and $g(\theta \!=\!0,\omega) =$ $g_{\rm o}$, where
$T_{\rm eff}^{\rm o}$ and $g_{\rm o}$ are the fundamental
parameters of the rotationless star, do not hold and that only
the effective temperature and gravity averaged over the whole
polar-on seen stellar hemisphere approach the effective
temperature $T_{\rm eff}^{\rm o}$ and gravity $g_{\rm o}$
roughly, i.e.: $\lim_{i\rightarrow 0}<\!\!T_{\rm
eff}(\theta,\omega|i)\!\!>$ $\sim T_{\rm eff}^{\rm o}$ and
$\lim_{i\rightarrow 0}<\!\!g(\theta,\omega|i)\!\!>$ $\sim g_{\rm
o}$. In all cases it is $T_{\rm eff}(\theta\!\!=\!\!0,\omega)
\geq$ $T_{\rm eff}^{\rm o}$ and $g(\theta \!\!=\!\!0,\omega)
\geq$ $g_{\rm o}$.\par


\subsection{Local plane-parallel models}
\label{sec:plan}

 The spectrum of a rotating star is represented as the intensity
of radiation emitted per unit wavelength interval per steradian
in the direction towards the observer defined by the aspect angle
i (inclination angle of the stellar rotation axis with respect to
the line of sight). It is given by \citep{1965ApJ...142..265C,
1970A&A.....7..120M}:

\begin{equation}
L_{\lambda}(i,\omega) = 2\int_0^{\pi/2}\!\!{\rm
d}\phi\!\int_0^{\pi}\!\!R^2(\theta)I_{\lambda}(\mu,\omega)
\frac{|\mu|}{\cos\delta}\sin\theta{\rm d}\theta
\label{eq:luminosity}
\end{equation}

\noindent where $R(\theta)$ is the co-latitude $\theta$-dependent
distance between the stellar center and the surface of the
rotation ally-distorted star;
$\mu=\mu(\phi,\omega)=\hat{n}.\hat{\i}$ ($\hat{n}$ is the unit
vector normal to the stellar surface and $\hat{\i}$ is the unit
vector representing the direction of the line of sight);
$\cos\delta(\theta,\omega)=-\hat{n}.\hat{r}$ ($\hat{r}$ is the
unit vector in the direction of $R(\theta)$);
$I_{\lambda}(\mu,\omega)$ is the $\mu$-dependent monochromatic
specific radiation intensity calculated for the local effective
temperatures $T_{\rm eff}(\theta,\omega)$ and surface gravities
$g(\theta,\omega)$. The integration of equation
(\ref{eq:luminosity}) was performed using a gauss-biquadrature of
degree 96 in both angles $\theta$ and $\phi$, which in principle
is equivalent to a grid of 36481 surface elements of local
plane-parallel model atmospheres. Each stellar surface point was
characterized by its specific radial velocity in the observer's
direction in order to take into account the corresponding Doppler
shift in the spectral line.\par
 The models we used to evaluate each local specific intensity
$I_{\lambda}(\mu,\omega)$ were computed in two consecutive steps.
To account in the most effective way for the line-blanketing, the
temperature structure of the atmospheres have been computed as in
Kurucz \& Castelli (2000) using the {\sc atlas}9 computer code
\citep{cdrom13}. Non-LTE level populations were then computed for
each of the atoms we considered using {\sc tlusty}
\citep{1995ApJ...439..875H} and keeping fixed the temperature and
density distributions obtained with ATLAS9.\par
 Excepted for C~{\sc ii}, the atomic models we used in this work
were downloaded from {\sc tlusty}'s homepage
(http://tlusty.gsfc.nasa.gov) maintained by I. Hubeny and T.
Lanz. Table \ref{tab:desc} lists the ions that were introduced in
our computations. C {\sc ii} was treated with use of the {\sc
modion} IDL package developed by \citet{modion} and by adopting
the atomic data (oscillator strengths, energy levels and
photoionization cross--sections) selected from the {\sc topbase}
database \citep{1993A&A...275L...5C}. It reproduces the results
obtained by \citet{1996ApJ...473..452S}.\par
 In this way and for each spectral region studied in the present
work, the specific intensity grids were computed for effective
temperatures and surface gravities ranging respectively from 15000
K to 27000 K and from 3.0 to 4.5. For \top\ $<$ 15000 K and \top\
$>$ 27000 K (\top\ concern rotationless model atmospheres) we
used LTE calculations and the {\sc ostar}2002 model atmospheres
grid \citep{2003ApJS..146..417L}.\par

\begin{table}[t!]
\caption{Atomic models used} \label{tab:desc}
\begin{tabular}{lll}
\hline
\noalign{\smallskip}
Atom & Ion & Description\\
\noalign{\smallskip}
\hline
\noalign{\smallskip}
Hydrogen & H {\sc i} & 8 levels + 1 superlevel\\
         & H {\sc ii} & 1 level\\
Helium   & He {\sc i} & 24 levels\\
         & He {\sc ii} & 20 levels\\
         & He {\sc iii} & 1 level\\
Carbon   & C {\sc ii} & 53 levels all individual levels\\
         & C {\sc iii} & 12 levels\\
         & C {\sc iv}  & 9 levels + 4 superlevels\\
         & C {\sc v} & 1 level\\
Nitrogen & N {\sc i} & 13 levels\\
         & N {\sc ii} & 35 levels + 14 superlevels\\
         & N {\sc iii} & 11 levels\\
         & N {\sc iv} & 1 level\\
Oxygen   & O {\sc i} & 14 levels + 8 superlevels\\
         & O {\sc ii} & 36 levels + 12 superlevels\\
         & O {\sc iii} & 9 levels\\
         & O {\sc iv} & 1 level\\
Magnesium & Mg {\sc ii} & 21 levels + 4 superlevels\\
          & Mg {\sc iii} & 1 level\\
\noalign{\smallskip}
\hline
\end{tabular}
\end{table}

\section{Comparison of rotating with non-rotating models}
\label{sec:comprnr}
\subsection{Adopted procedure}
\label{sec:proc}

\begin{figure*}
\center
\includegraphics[width=16cm,angle=0,clip=]{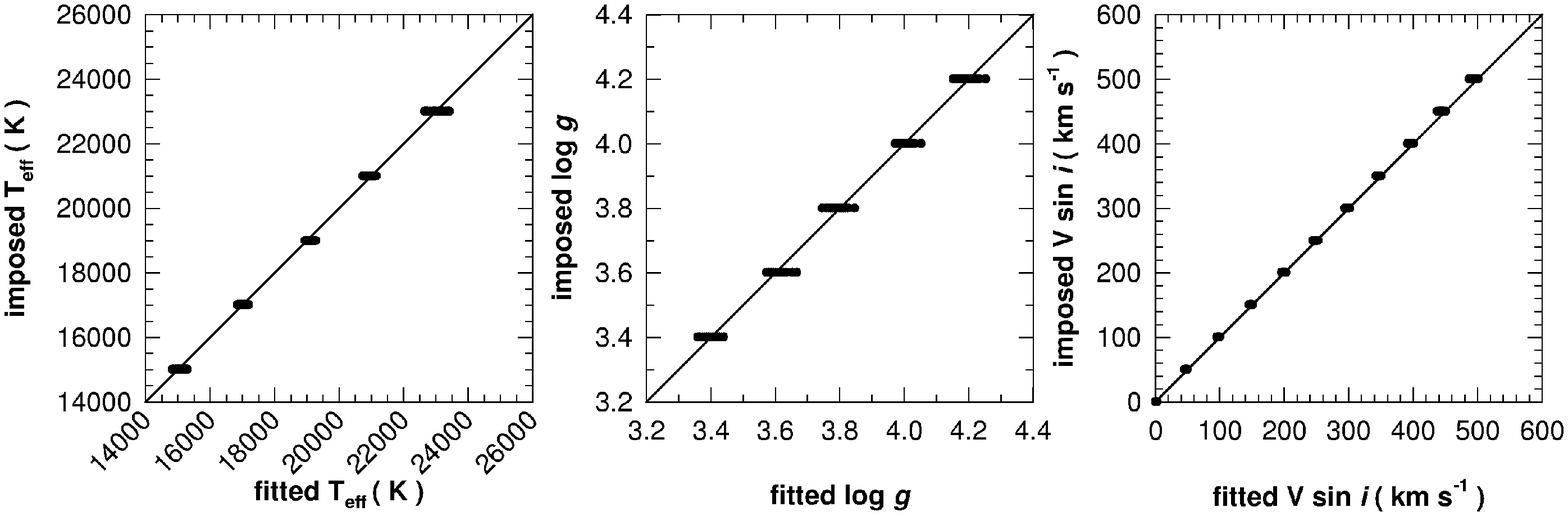}
\caption{Test of the adopted procedure: comparison of
$T_{\rm eff}$, $\log g$ and \vsini\ parameters obtained by the
fitting procedure with the model parameters of GRIDA}
\label{test}
\end{figure*}

 Hydrogen and helium line profiles, more particularly, the H$\gamma$,
$\lambda$4388 \ion{He}{i} and $\lambda$4471 \ion{He}{i} lines, are
often used to derive the effective temperature, surface gravity and
projected rotation velocity of B type stars. Among the reasons of
their use, we have: 1) the line profiles and equivalent widths are
very sensitive to effective temperature and surface gravity; 2) the
blue helium lines are known to be somewhat less sensitive to LTE
departures; 3) the broadening mechanisms of their line profiles have
been well known for a long time \citep[e.g.:][]{1969A&A.....1...28B,
1971A&A....11..387L,1973ApJS...25...37V,1974ApJ...190..315M}.\par
 Two grids of spectra ranging from 4250 \AA~to 4500 \AA~were
computed. A first grid, GRIDA, was obtained using classical
plane-parallel model atmospheres. A second grid, GRIDB, was
calculated using {\sc fastrot} (Sect.~\ref{sec:ma}) where stellar
flattening and gravity darkening are taken into account. The sets
of fundamental parameters considered in these grids are given in
Table \ref{sets}. $T_{\rm eff}^{\rm o}$ and $g_{\rm o}$ refer to
the fundamental parameters of the rotationless or parent
non-rotating counterpart of the star (Sect.~\ref{sec:flat}).
Regarding the notation, it is worth noting that for GRIDA it
holds $T_{\rm{eff}} =$ $T_{\rm eff}^{\rm o}$ and $\log g =$ $\log
g_{\rm o}$.\par

\begin{table}[t]
\caption{Sets of parameters used in model grids}
\label{sets}
\begin{tabular}{@{}rcll@{}}
\hline
\noalign{\smallskip}
{ GRIDA} & & & \\
\hline \noalign{\smallskip}
 5000 K~~$\le$ & \top &$\le$~~40000 K    & step = 1000 K  \\
     2.5~~$\le$ & \gop      &$\le$~~5.0        & step = 0.5     \\
  0 \kps~~$\le$ & \vsini\       &$\le$~~500 \kps   & step = 50 \kps \\
\noalign{\smallskip}
\hline
\noalign{\smallskip}
{ GRIDB} & & & \\
\hline\noalign{\smallskip}
12000 K~~$\le$  &$T_{\rm eff}^{\rm o}  $&$\le$~~27000 K& step = 1000 K\\
    3.2~~$\le$  &$\log g_{\rm o}       $&$\le$~~4.4    & step = 0.2   \\
0$^{\rm o}~~\le$&$i$          &$\le$~~90$^{\rm o}$     & step = 5$^{\rm o}$\\
     0.50~~$\le$&$\Omega/\Omega_{\rm c}$&$\le$~~0.99   & step = 0.05  \\
\noalign{\smallskip}
\hline
\end{tabular}
\end{table}

 In what follows we will determine the effective temperature,
surface gravity and the projected rotational velocity of B-type
stars using only the H$\gamma$, He{\sc i}4388, He{\sc i}4471 and
Mg{\sc ii}4481 lines. To this end we will use a fitting procedure
based on a least-squares method that applies the {\sc minuit}
minimization package available at {\sc cern}. In order to know
the confidence degree of this method, we fitted the spectra of
the GRIDA reference lybrary using synthetic spectra obtained from
the same NLTE plane-parallel model atmospheres like the GRIDA
spectra. The free, or fit parameters used to this purpose were:
the effective temperature, surface gravity, the projected
rotational velocity and the averaged flux ratio of the fitted to
the reference spectrum. The $\chi^2$ deviation parameter was
computed only for the selected spectral domains defined in Table
\ref{tab:fitr} for CASE A. The comparison shown in
Fig.~\ref{test} of parameters obtained by the minimization
procedure with the model GRIDA reference parameters, reveals that
the agreement is quite satisfactory. However, this agreement is
the best for the projected rotational velocity, whose accuracy is
generally better than 4\%. The derived effective temperature
approaches the most to the reference values when $T_{\rm eff}
\sim$ 20000 K, but for other temperatures deviations never exceed
500 K. The errors committed on $\log g$ are always lower than
0.05 dex whatever the effective temperature.\par

\begin{table}[t]
\center
\begin{minipage}{8cm}
\caption{Spectral regions considered for the fits.}
\label{tab:fitr}
\begin{tabular}{crcc}
\noalign{\smallskip}
\hline
\noalign{\smallskip}
     &\multicolumn{2}{c}{CASE} & \\
 Spectral line & A & B & $\Delta\lambda$ (\AA) \\
\noalign{\smallskip}
\hline
\noalign{\smallskip}
 H$\gamma$ & {\large $\times$} & {\large $\times$} & 4300 - 4375 \\
$\lambda$4388 \ion{He}{i} & {\large $\times$} & {\large $\times$}  & 4375 - 4400 \\
$\lambda$4471 \ion{He}{i} &   & {\large $\times$}  & 4460 - 4475 \\
$\lambda$4471 \ion{He}{i} + $\lambda$4481 \ion{Mg}{ii} & {\large $\times$} & &  4460 - 4490 \\
\noalign{\smallskip}
\hline
\end{tabular}
\end{minipage}
\end{table}

 To study the effects of fast rotation on the determination of
stellar fundamental parameters, we applied the same procedure
than above to the GRIDB spectra, where the model parameters are
$\{T_{\rm eff}^{\rm o},\log g_{\rm o},i,$\omc$\}$ or equivalently
$\{T_{\rm eff}^{\rm o},\log g_{\rm o}$,\vsini,\omc$\}$. Since the
fit of spectra produced by rotating stars can be done in two
ways, we distinguish two sets of parameters. The reference model
parameters proper to calculate {\sc fastrot} rotationally
modified synthetic spectra, as well as those obtained by fitting
with them the rotationally distorted spectra, are hereafter
called { {\it parent non-rotating counterpart (pnrc)}}
fundamental parameters { and will be noted with a superscript
``~$^{\rm o}$~'' or a subscript ``~$_{\rm o}$~'' (i.e. \top\ and
\gop), while \vsini\ will represent the true projected rotation
velocity of the star}. Conversely, we call {\it apparent}
fundamental parameters { (i.e. \tap, \gap\ and \vsiniap)}, those
derived by fitting the GRIDB spectral lines with synthetic
spectra issued from classical plane-parallel model
atmospheres.\par The fit of the H$\gamma$, He{\sc i}4388 and
He{\sc i}4471 lines was done either including the Mg{\sc ii}4481
line or excluding { it in order to test the procedure's
sensitivity to a change of the fitting criteria; this identifies
respectively CASE A or CASE B in Table~\ref{tab:fitr}.}\par

\begin{figure*}[ht!]
\center
\includegraphics[width=18cm,angle=0,clip=]{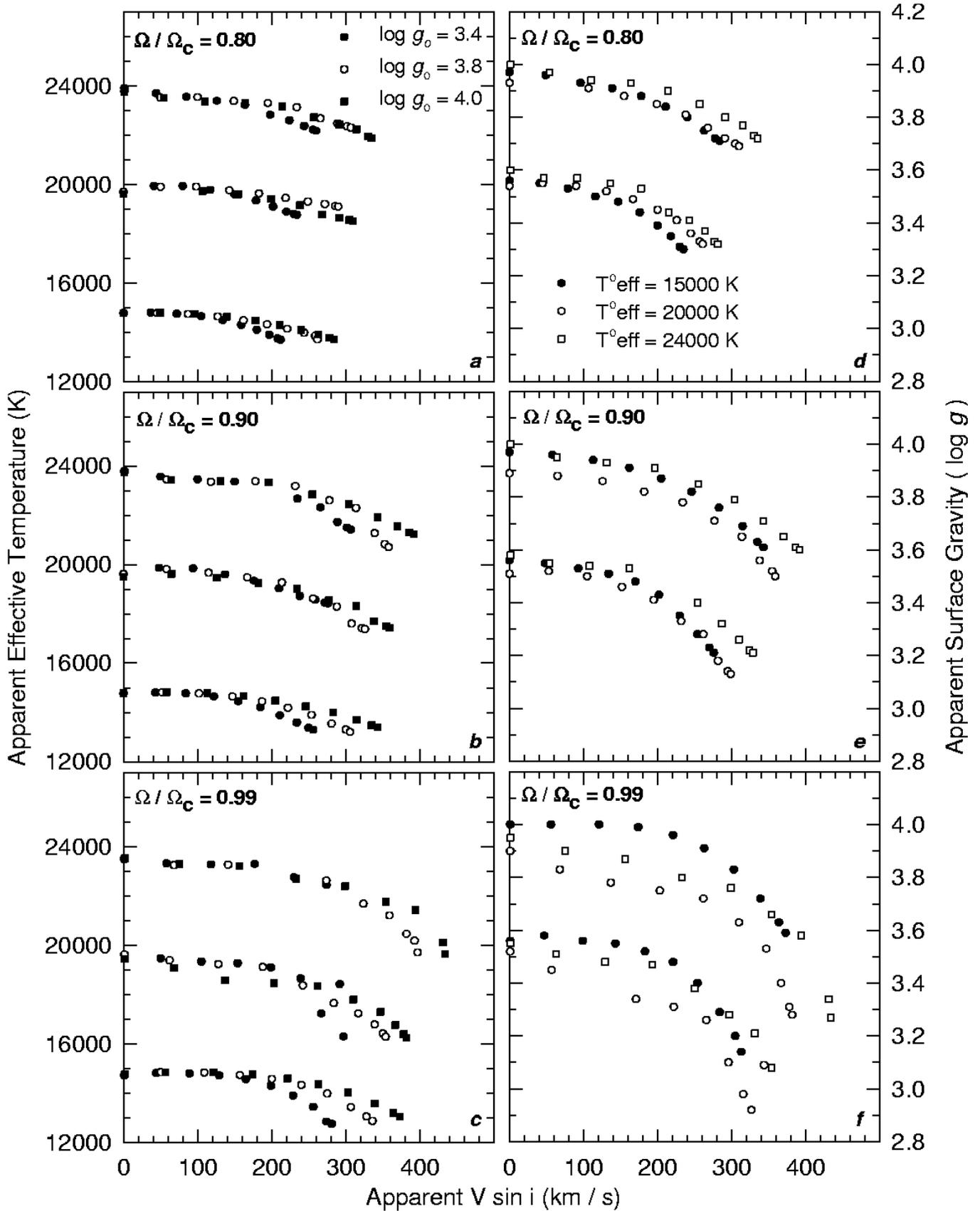}
\caption{Apparent effective temperature and surface gravity as a
function of \vsini\ for different pnrc (\top,\gop) values and
rotation rates \omc} \label{fig:otemp}
\end{figure*}

\subsection{Reference parameters}

 In the following sections, calculations will be done for several
effective temperatures and surface gravities. To have a rough look
on the spectral types they may concern as well as the relations
that exist between the fundamental parameters suited for
non-rotating stars and the apparent ones -- which are dependent on
the stellar geometrical distortions and the concomitant GD -- we
list in Table \ref{refparam} the reference parameters we used in
the present work. In this table are also given the polar
effective temperature and the polar radius of the distorted stars
for several values of the angular velocity ratio \omc.

The MK spectral types adopted \citep{1994AJ....107..742G} are for
the effective temperatures and \gop~= 4.0 of the { {\it parent}}
non-rotating stars. The stellar masses and radii for \omc\ = 0
used to calculate the critical velocity $V_{\rm c}$ (c.f Sect.
\ref{sec:flat}) were derived from stellar evolutionary tracks of
non-rotating stars \citep{1992A&AS...96..269S}.\par

\begin{table}[t!]
\caption{Polar temperature (\tpole), polar radii (\rpole) and
critical velocities ($V_{\rm{c}}$) as a function of \omc\ and for
\gop = 4.} \label{refparam}
\begin{tabular}{ccccccc}
\noalign{\smallskip} \hline \noalign{\smallskip} MK Type &
\top     & Mass         & \omc & \tpole     & \rpole &
$V_{\rm{c}}$\\
        &     K   & M$\odot$   &      &     K    & R$\odot$  & km/s \\
\noalign{\smallskip}
\hline
\noalign{\smallskip}
 B1 V  & 24620 & 10.53 & 0.00 & 24620  &  5.32  & 525 \\
       &       &       & 0.50 & 24932  &  5.22  &     \\
       &       &       & 0.80 & 25887  &  5.10  &     \\
       &       &       & 0.99 & 27593  &  4.88  &     \\
 B2 V  & 19500 & 6.87  & 0.00 & 19500  &  4.30  & 472 \\
       &       &       & 0.50 & 19736  &  4.21  &     \\
       &       &       & 0.80 & 20471  &  4.11  &     \\
       &       &       & 0.99 & 21773  &  3.94  &     \\
 B5 V  & 14000 & 4.15  & 0.00 & 14000  &  3.34  & 418 \\
       &       &       & 0.50 & 14176  &  3.27  &     \\
       &       &       & 0.80 & 14708  &  3.19  &     \\
       &       &       & 0.99 & 15609  &  3.04  &     \\
 B9 V  & 10720 & 2.86  & 0.00 & 10720  &  2.77  & 383 \\
       &       &       & 0.50 & 10870  &  2.71  &     \\
       &       &       & 0.80 & 11300  &  2.63  &     \\
       &       &       & 0.99 & 12007  &  2.50  &     \\
\noalign{\smallskip}
\hline
\end{tabular}
\end{table}

\section{Apparent fundamental parameters}

 Following the procedure described in Section \ref{sec:proc}, we
obtained a grid of {\it pnrc} (i.e.: parameters belonging to the
parent non-rotating stellar counter-part) and {\it apparent}
fundamental parameters for different values of the angular
velocity and inclination angle of the star. { Since the aim of
the present computations is to provide a way to correct the
observed fundamental parameters from gravitational darkening
effects for a large parameter-space, in the present work we have
chosen to present the effects due to fast rotation and the
corresponding corrections, as a function of \tap, \gap\ and
\vsiniap\ (e.g. Fig. \ref{fig:otemp}), to make a direct link with
the actually measured quantities, rather than against the pnrc
fundamental parameters, which need to be determined from the
sought corrections. As the REFs are strongly dependent on the
aspect angle $i$, we present them as a function of the observed
\vsiniap.}


\begin{table}[t!]
\center
\begin{minipage}{8cm}
\caption{Effects of fast rotation in equator-on B type stars. The
effective temperature (R\teff) and surface gravity (R\logg)
underestimation percentage are given against
\omc.}\label{tab:efrp} \center
\begin{tabular}{crc}
\hline
\omc & R\teff & R\logg\\
\noalign{\smallskip}
\hline
\noalign{\smallskip}
0.80 & 9~\% & 8~\%\\
0.90 & 12~\% & 10~\%\\
0.99 & 17~\% & 20~\%\\
\noalign{\smallskip}
\hline
\end{tabular}
\end{minipage}
\end{table}


\begin{figure*}
\center
\includegraphics[width=14.5cm,angle=0,clip=]{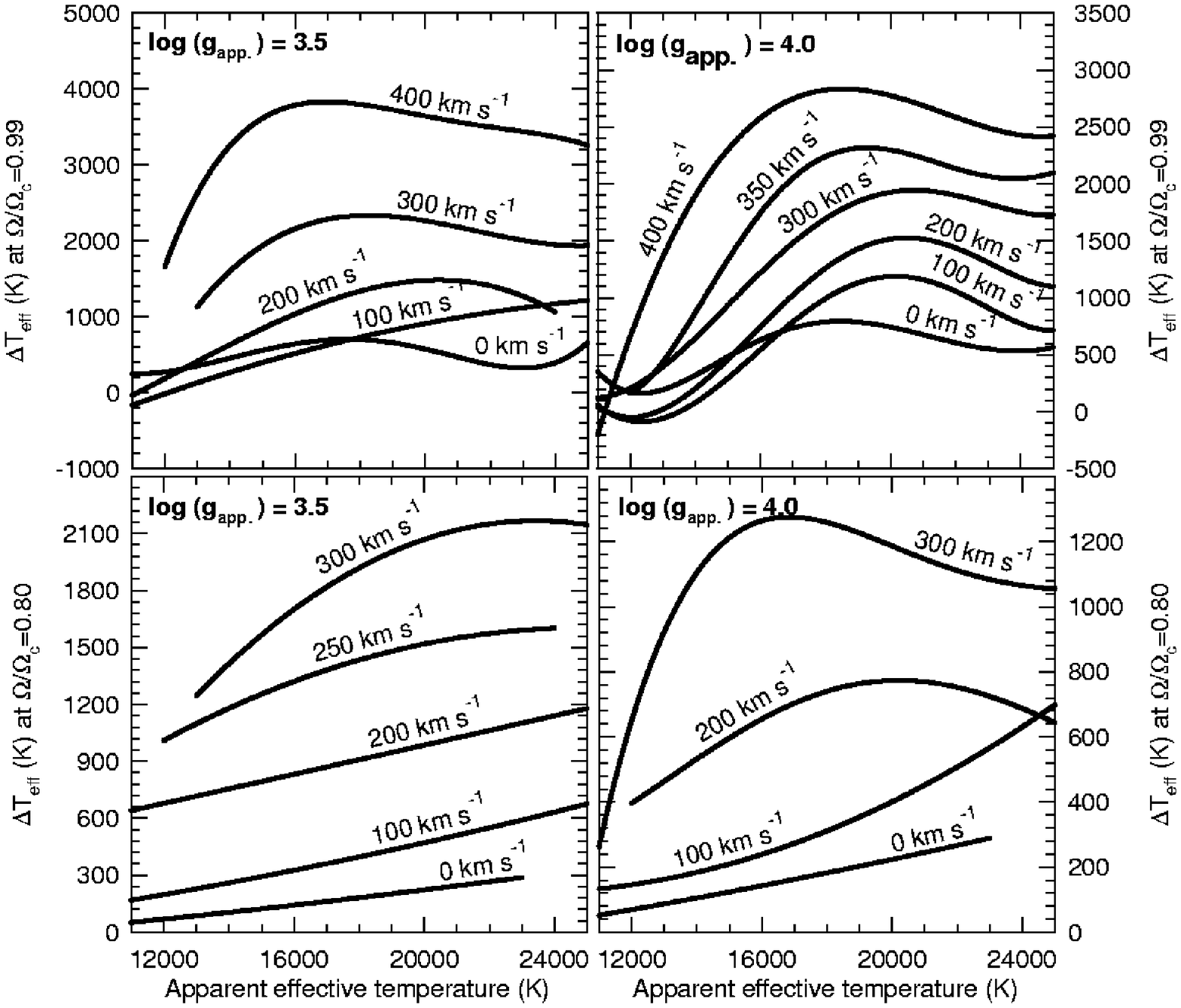}\\
\includegraphics[width=14.cm,angle=0,clip=]{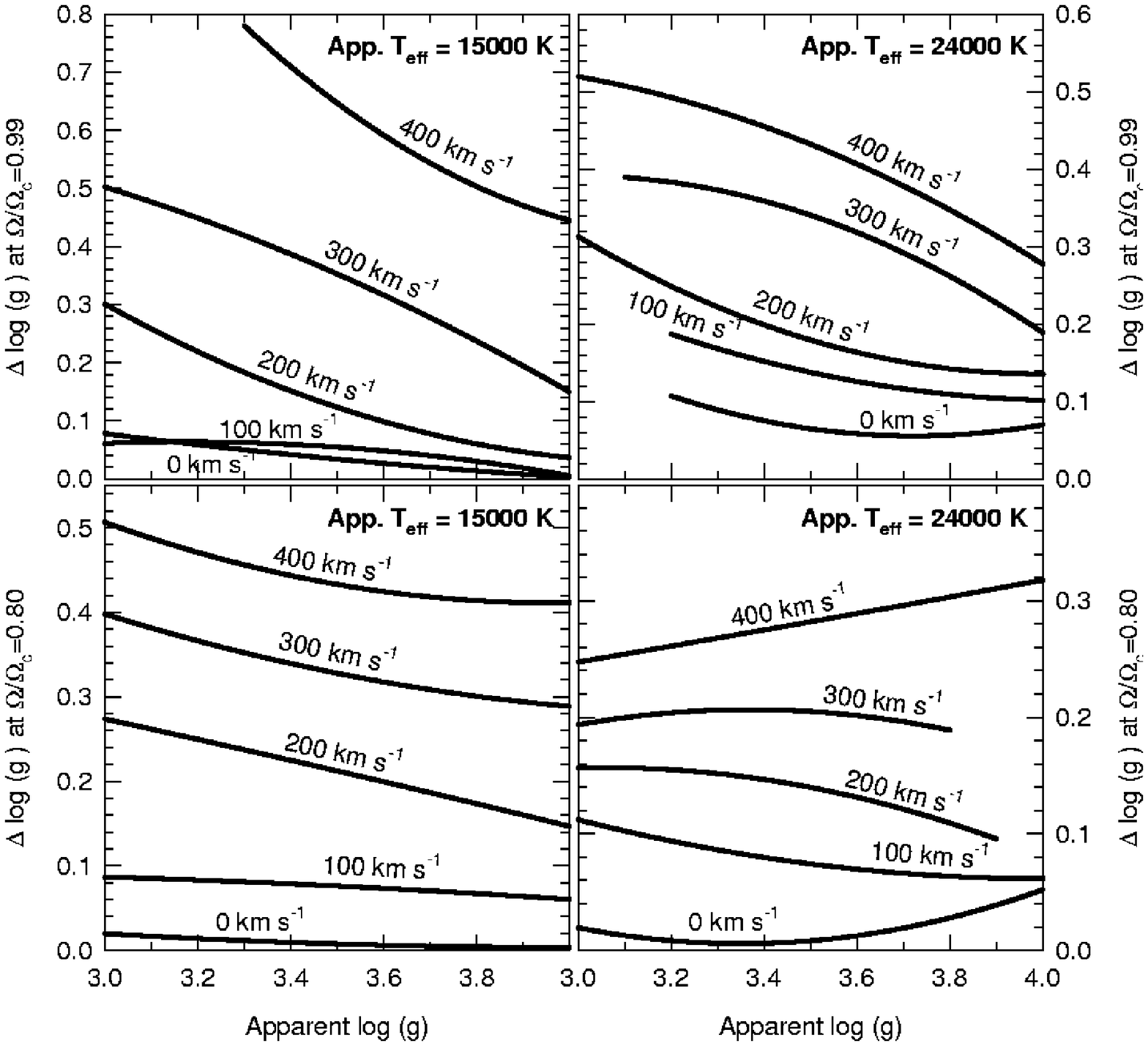}
\caption{Upper panel: Corrections $\Delta$\teff to add to the
apparent effective temperature (\tap) for different \omc,
\vsiniap, and \gap. Lower panel: Corrections $\Delta$\logg\ to
add to the apparent surface gravity for different \omc, apparent
\vsini, and apparent \teff} \label{fig:toutg}\label{fig:toutt}
\end{figure*}



\subsection{Effective temperature}

The typical effects of stellar flattening and GD on effective
temperature are shown in Fig. \ref{fig:otemp} for different
combinations of \top~and \gop, while the corrections
$\Delta$\teff\ to add to the {\it apparent} effective temperature
to get the pnrc effective temperature at different apparent \omc,
apparent \vsini, and apparent \logg\ regimes are plotted in
Fig.~\ref{fig:toutt} { (upper panel)}. As expected, the \tap\
decreases with increasing \vsiniap. Fast rotation produces
therefore a lowering of the apparent effective temperature.
Concerning the amplitude of this effect and its general
behaviour, the following remarks can be made:

\begin{enumerate}
\item[{\it i})] REFs on \tap\ become significant for \omc\
$>$ 0.6. They are stronger and vary more rapidly with \vsiniap\
the larger the values of $\Omega$.
\item[{\it ii})] When \omc\ increases, REFs on \tap\ become
more sensitive to surface gravity in the sense that they are
larger the lower is the gravity.
\item[{\it iii})] The relative lowering effect on the apparent
\teff~does not depend much on its pnrc value. In Table
\ref{tab:efrp} this effect is given in percentages for different
angular velocities in stars seen equator-on, where effects are
the stronger.
\item[{\it iv})] When studying dwarf B-type stars, REFs on \tap\
can be ignored safely at \vsiniap\ lower than 200 \kps\ and at
\vsiniap\ $\la$ 100 \kps\ for giant B-type stars.
\end{enumerate}

\subsection{Surface gravity}

Corrections $\Delta$\logg\ to add to the {\it apparent} surface
gravity to get the pnrc surface gravity at different \omc,
apparent \vsini\ and apparent \teff~regimes are plotted in
Fig.~\ref{fig:toutg} { (lower panel)}. As shown in Fig.
\ref{fig:otemp}, stellar flattening and gravitational darkening
produce a lowering of the apparent surface gravity. The amplitude
of this REF varies obeying the following rules:
\begin{enumerate}
\item[{\it i})] REFs on \gap\ become significant for \omc\ $>$0.6.
They are stronger and vary more rapidly with \vsiniap\ at high
\omc values.
\item[{\it ii})] When \omc\ $\la$ 0.85, REFs on \gap\ are weaker
in stars with \top\ $\ga$ 20000~K than in those with \top\ $\la$
20000~K. On the contrary, for \omc\ $\ga$ 0.85, the apparent
\logg\ are more sensitive to REFs in stars with \top\ $\ga$
20000~K than in those with \top\ $\la$ 20000~K. This change in the
\logg\ sensitivity to REFs, according to \top~and \omc, is due to
hydrogen lines, which in the present approach are the main \logg\
indicators. In hot stars, the intensity of hydrogen lines is {
the lowest}, but it increases fast with the equatorial local
\teff\ decrease produced by increasing values of \omc.
\item[{\it iii})] The lowering of \gap~does not depend much on the
pnrc \logg~value. It is given in percentages in Table
\ref{tab:efrp} for different angular velocities in stars seen
equator-on where effects are the stronger. In the more distorted
cases, the change in \gap~represents two luminosity classes.
\item[{\it iv})] In B-type stars with \tap\ $\la$ 18000~K and
\vsiniap\ $\la$ 100~\kps, REFs on \logg~can be ignored.
\end{enumerate}

\subsection{Projected rotation velocity}
\label{sec:prv}

\begin{figure}
\center
\includegraphics[width=7cm,angle=0,clip=]{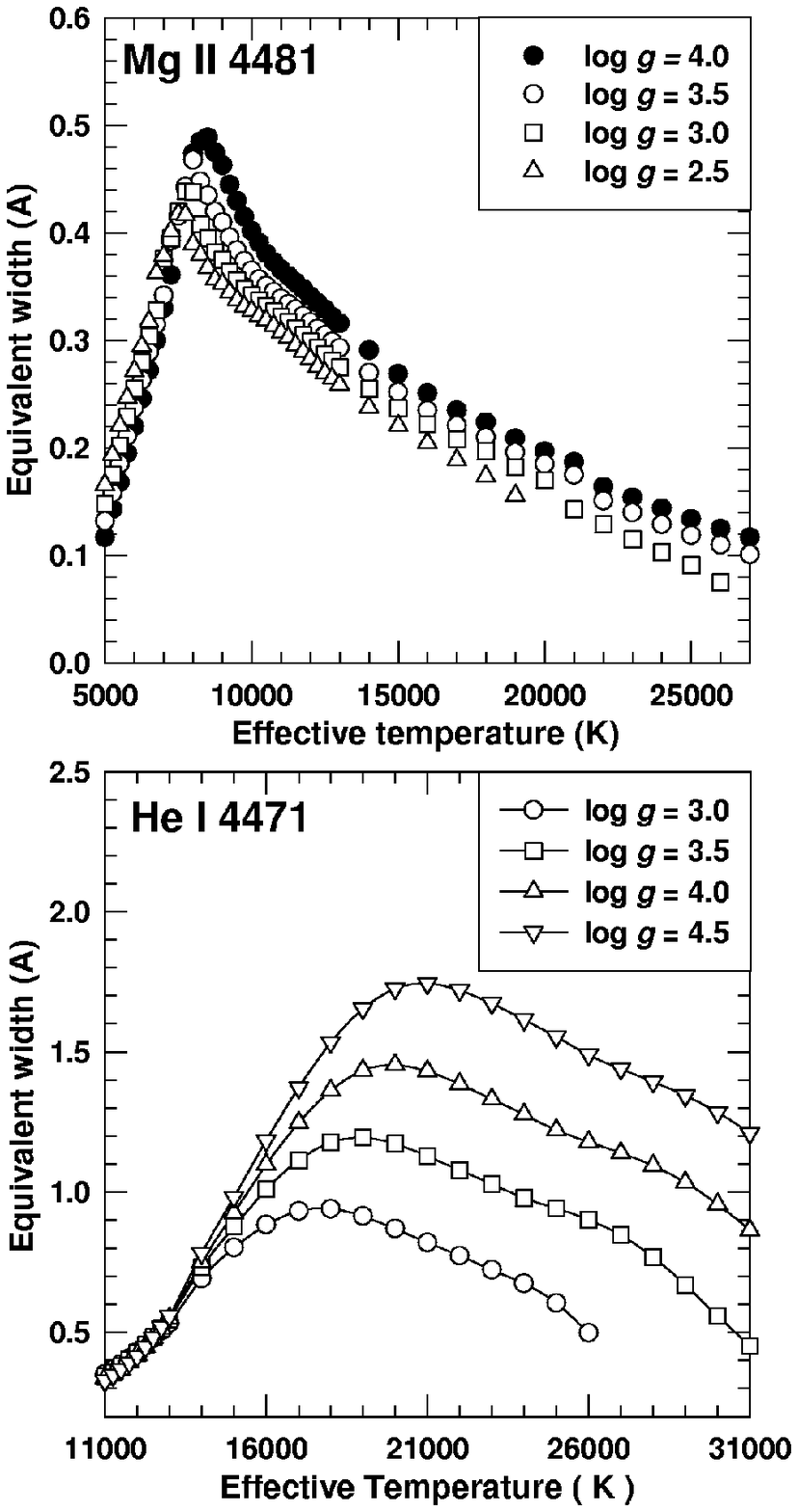}
\caption{Equivalent widths of the \ion{He}{i} 4481 and
\ion{Mg}{ii} 4481 spectral lines against effective temperature
for different \logg~values. Computations were made assuming
plane-parallel model atmospheres.} \label{fig:twl}
\end{figure}

The REFs on \vsiniap~become significant for \omc\ $\ga$ 0.70. The
spectral lines become progressively less broadened than expected
from models of stars without GD, due to a lower contribution to
the line flux coming from the gravity darkened equatorial regions
towards the stellar limb. The magnitude of this effect depends on
the sensitivity of the line studied to the local changes of
\teff~and~\logg, as shown in Fig. \ref{fig:twl} for the
\ion{He}{i}~4471 and for the \ion{Mg}{ii}~4481 lines.\par

\begin{figure}[t]
\center
\includegraphics[width=7cm,angle=0,clip=]{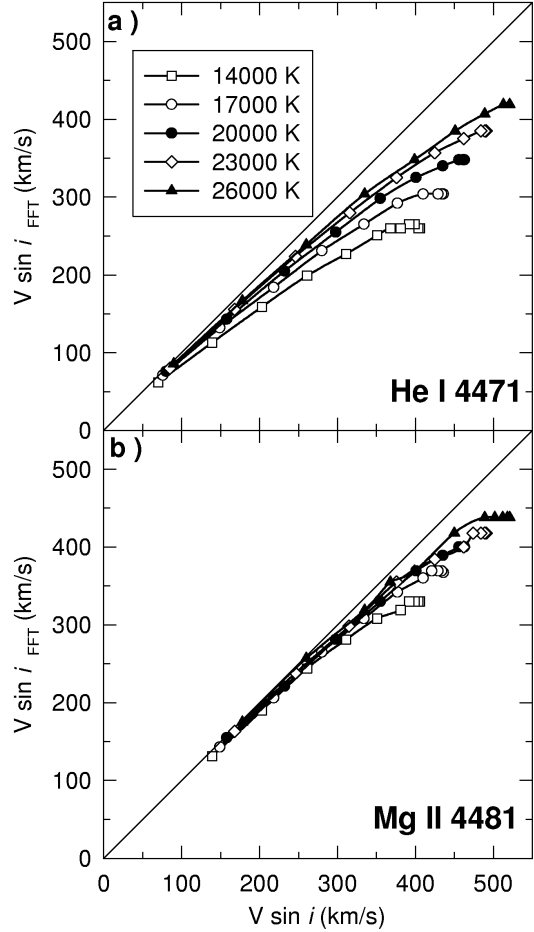}
\caption{Apparent \vsini\ derived using the Fourier Transform
technique against the true \vsini\ considering individually the
\ion{He}{i}~4471 line (panel a) and \ion{Mg}{ii}~4481 spectral
line (panel b). Computations are made assuming \omc\ = 0.99 .}
\label{fig:fftvsini}
\end{figure}

\begin{figure}[t]
\includegraphics[width=7cm,angle=0,clip=]{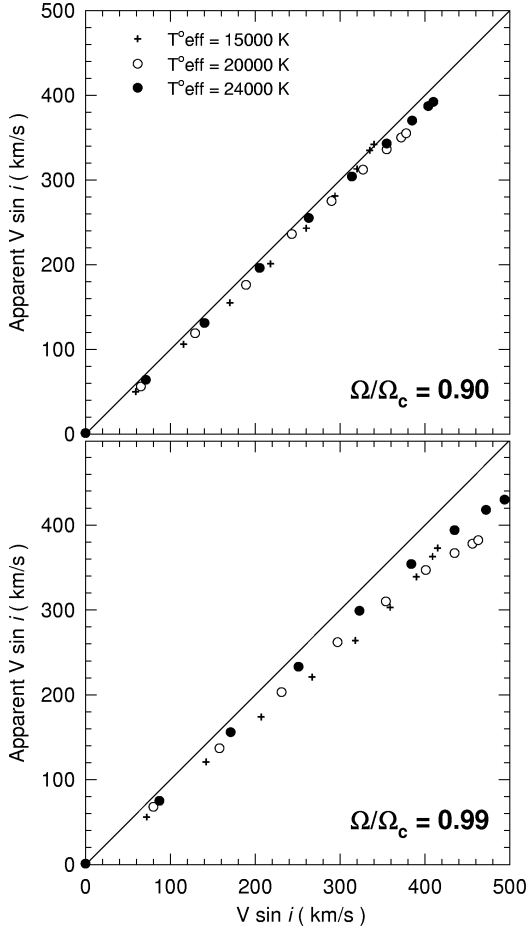}
{\center\caption{Comparison between true and apparent \vsini~in
fast rotating stars assuming \gap = 4. The fitted regions
correspond to CASE A of Table~\ref{tab:fitr}.}\label{fig:vst}}
\end{figure}

\begin{figure}[t]
\center
\includegraphics[width=8.7cm,angle=0,clip=]{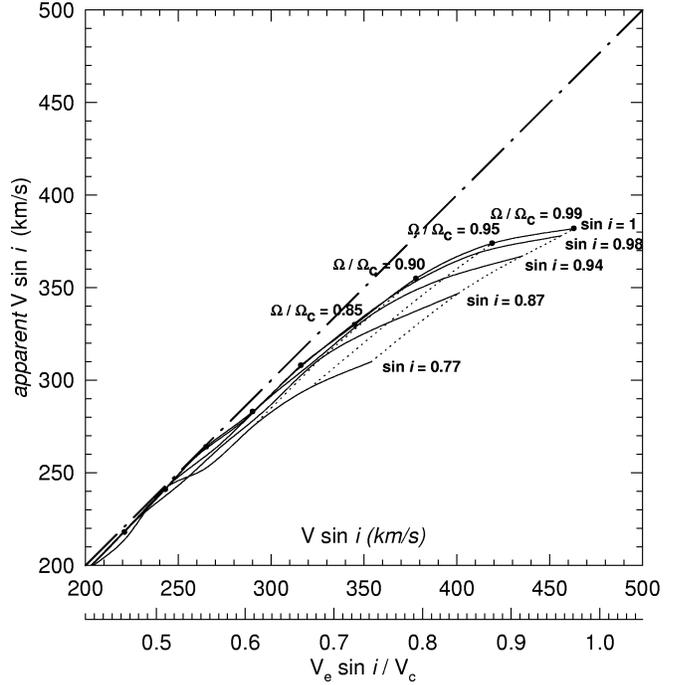}
\caption{{ Apparent and true \vsini\ values at \top = 20000~K,
\gop = 4.0 and different values of $\sin~i$ (full-line curves) and
of \omc\ (dotted curves).} }\label{fig:fitr}
\end{figure}

\begin{figure}[t]
\center
\includegraphics[width=7cm,height=14cm,angle=0,clip=]{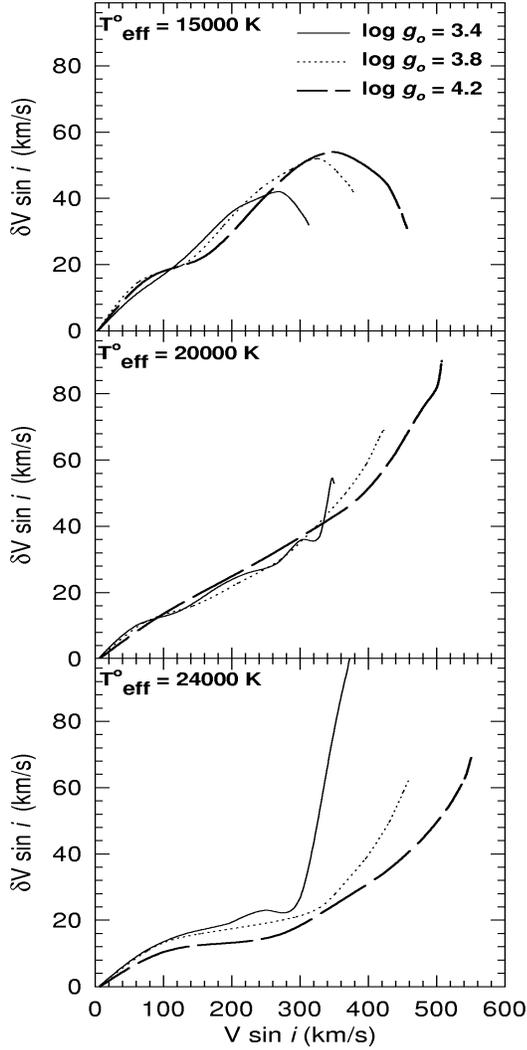}
\caption{\vsini~lowering ($\delta\!V\!\sin i$) as a function of
the true \vsini~for different true effective temperature and true
\logg~values.} \label{fig:difef}
\end{figure}

 At a given \omc\ $\ga$ 0.7 and \top\ $\la$ 18000~K, the
broadening of the \ion{He}{i}~4471 line tends to saturate for
increasing values of \vsiniap. In stars cooler than \top\ $\simeq$
18000~K, the saturation of this broadening is favored also by the
enhanced contribution to the line flux from the polar regions,
where the local \teff\ is increased by the GD.
Fig.~\ref{fig:fftvsini} shows the apparent \vsini$_{\rm
FFT}$~against true \vsini\ for \gop\ = 4.0, \omc\ = 0.99 and for
different values of \top\ when only the \ion{He}{i}~4471 line
(Fig.~\ref{fig:fftvsini}~a) is considered. The apparent
\vsini~parameter is obtained in this figure using the Fourier
transform method applied to line profiles calculated with {\sc
fastrot}. In the present work, the Fourier method is meant to
represent the classical techniques of determining rotational
parameters. We see that in a star having a \top\ = 14000~K, a
\vsiniap\ $\simeq$ 250~\kps\ value implies that the
underestimation of the projected rotational velocity is
$\delta$\vsini = (\vsini)-(\vsiniap)$\simeq 140$~\kps, if it
rotates at \omc\ =0.99. The underestimation $\delta$\vsini\ drops
fast as soon as \omc\ $<$ 0.99, { while the magnitude of the
\vsiniap\ saturation depends on the inclination angle, angular
velocity and also on the effective temperature (e.g.
Fig.~\ref{fig:fftvsini}, \ref{fig:vst}, \ref{fig:difef} and
\ref{fig:fitr})}.\par

{ Same conclusions can be drawn for the \ion{Mg}{ii} 4481 line.
However, in this case the above mentioned line broadening
underestimation is less pronounced (Fig.~\ref{fig:fftvsini}~b).}
This is due to the fact that the intensity of the \ion{Mg}{ii}
4481 line is smaller in the polar regions than in the equatorial,
so that even the continuum intensity in the line wavelengths is
stronger at the pole than in the equator. The differentiated
response of the continuum and the intrinsic line intensity to
\teff~and \logg~compensate to each other somewhat, producing
hence a smaller \dvsini.

The relation between the apparent and the true \vsini~derived
from the \ion{Mg}{ii}~4481 is shown in Fig.~\ref{fig:fftvsini}.
We note in this figure that saturation effects are seen in the
\ion{He}{i}~4471 line for low \top\ values and in the
\ion{Mg}{ii} 4481 line when \top\ is high. In the first case, the
effect is due to a significant decrease of the equivalent width
of the line for low effective temperatures { (i.e. at the stellar
equator)}. In the second case, the contrast of the continuum
specific intensity between the pole and the equator cannot be
compensated by the modest increase of the equivalent width of the
line at lower equatorial temperatures (cf. Fig.
\ref{fig:twl}).\par

{ As can be seen in Fig. \ref{fig:vst}, where \vsiniap\ was
derived following the procedure described in Sect.~3.1, the
simultaneous fitting of various spectral lines combined with the
simultaneous determination of \tap, \gap\ and \vsiniap\ limits
somewhat the saturation effect and minimizes the underestimation
of the projected rotation velocity at high inclinations and
angular velocity rates.} Fig.~\ref{fig:vst} shows the relation
between the apparent and true \vsini~for \omc\ = 0.80, \omc\ =
0.99, \gop~= 4.0 and for different \top\ values when both
\ion{He}{i} 4471 and \ion{Mg}{ii} 4481 lines are taken into
account for the \vsini\ determination. Considering the case \omc\
= 0.99 { (i.e. \vevc\ $\simeq$ 0.97)}, we see that for whatever
pnrc \teff, the apparent \vsini\ underestimates the true rotation
by \dvsini\ $\la$ 50~\kps\ as far as the apparent \vsini\ $\la$
350~\kps. Underestimations may then increase to \dvsini\ $\simeq$
80~\kps for apparent \vsini\ $\ga$ 350~\kps. It is worth noting
the peculiar behavior of this relation for the \top\ = 15000~K.
The enhanced contribution of the \ion{Mg}{ii}4481 line to the
blend \ion{He}{i}4471+\ion{Mg}{ii}4481 at high inclinations,
makes that the \vsini\ is likely determined by the behavior of the
\ion{Mg}{ii}4481 which, as seen in Fig.~\ref{fig:fftvsini}, leads
to a smaller \dvsini\ than the \ion{He}{i} line alone. It is then
crucial that for stars with \top\ $\la$ 15000~K both
\ion{He}{i}4471 and \ion{Mg}{ii}4481 are considered together in
the \vsini\ calculation. On the contrary, for \top\ $\ga$ 20000~K,
the addition of $\lambda 4481$ \ion{Mg}{ii} in the fit does not
increase significantly the accuracy of the \vsini\ measurements.
Fig. \ref{fig:vst} also shows that for high enough \omc\ ratios
the underestimation \dvsini~is a function of \top.\par

{ As the use of the \vsini\ parameter at a given \omc\ hides the
effect of the aspect angle on the measured \vsini\, we plotted in
Fig.~\ref{fig:fitr}, for a pnrc B2~V spectral type, the behaviour
of the apparent \vsini\ against its true value at different fixed
inclination angles ($\sin i$). We note that the underestimation
at higher \vsini\ is also very sensitive to the $\sin i$.}\par

The incidence of \gop\ on the underestimation \dvsini\ is {
further} shown in Fig.~\ref{fig:difef}. In this figure the
relation between $\delta\!V\!\sin i$ against the true \vsini\ is
given for different pnrc effective temperatures and surface
gravities. It is seen that as long as \vsini\ $\la$ 300 \kps, the
$\delta\!V\!\sin i$ is almost independent of gravity. As soon as
\vsini\ $\ga$ 300 \kps, not only do the $\delta\!V\!\sin i$
differences depend strongly on \gop, but depending on the
effective temperature it shows two different behaviors. For
temperatures \top\ $\la$ 20000 K, $\delta\!V\!\sin i$ begins to
decrease, while for \top\ $\ga$ 20000 K it increases even faster.
In Fig.~\ref{fig:difef} the curve corresponding to \gop\ = 3.4 is
badly determined for high \vsini\ values, because { the gravity
decreases rapidly to very low values at the equator.}\par

\section{Effects of fast rotation in other spectral regions}

\begin{figure*}[ht!]
\center
\begin{tabular}{c}
\includegraphics[width=15cm,height=11.35cm,angle=0,clip=]{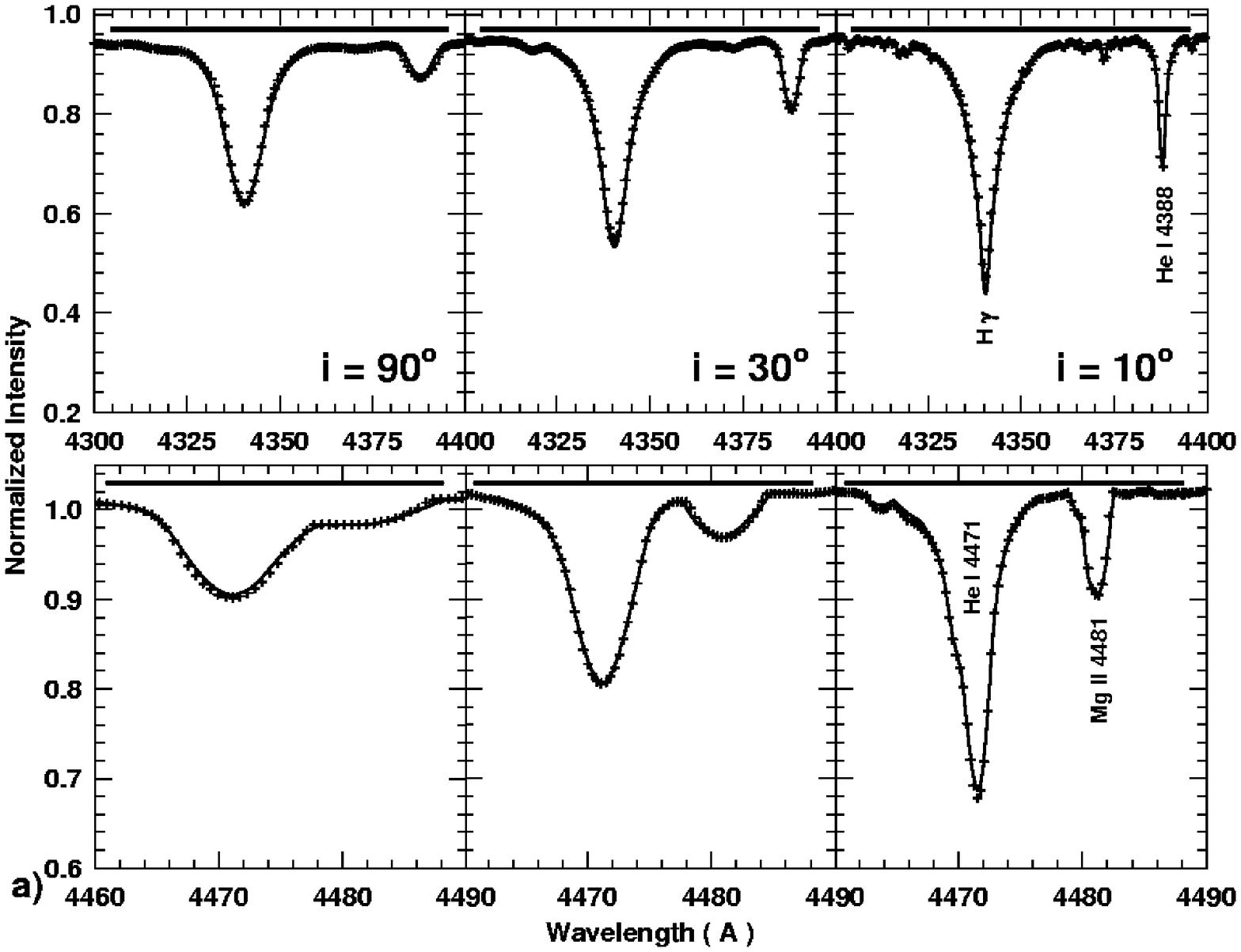}\\
\includegraphics[width=15cm,height=11.35cm,angle=0,clip=]{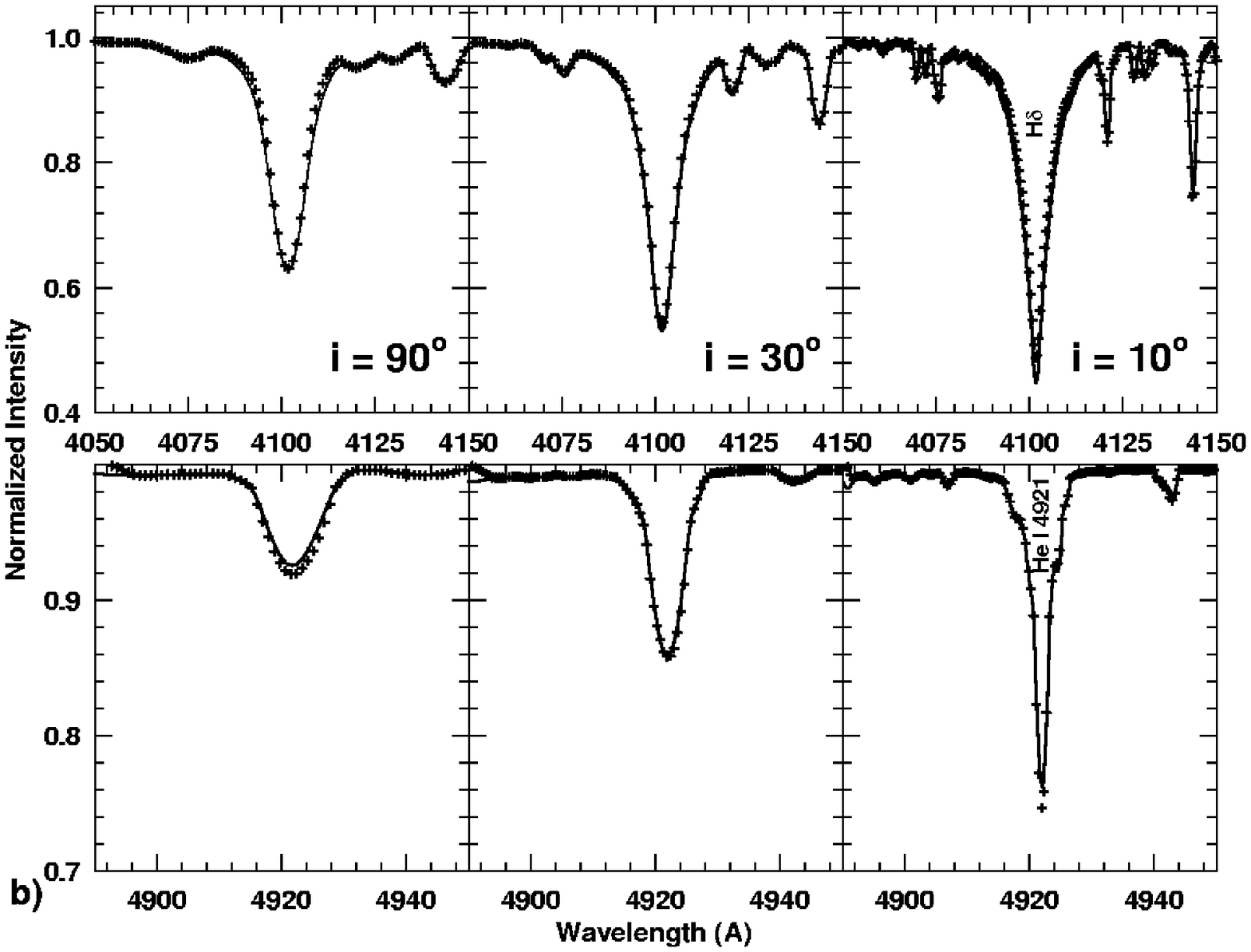}
\end{tabular}
\caption{Comparison between spectra computed using pnrc (full
line) and apparent parameters (crosses). Pnrc spectra were
computed for \top~=~20000~K, \gop~=~4.0 and \omc~=~0.99. Three
different inclination angles were considered. The values of the
apparent parameters for each case are given in
Table~\ref{tab:valap}.} \label{fig:r1}
\end{figure*}

\begin{figure*}[ht!]
\center
\begin{tabular}{c}
\includegraphics[width=16cm,angle=0,clip=]{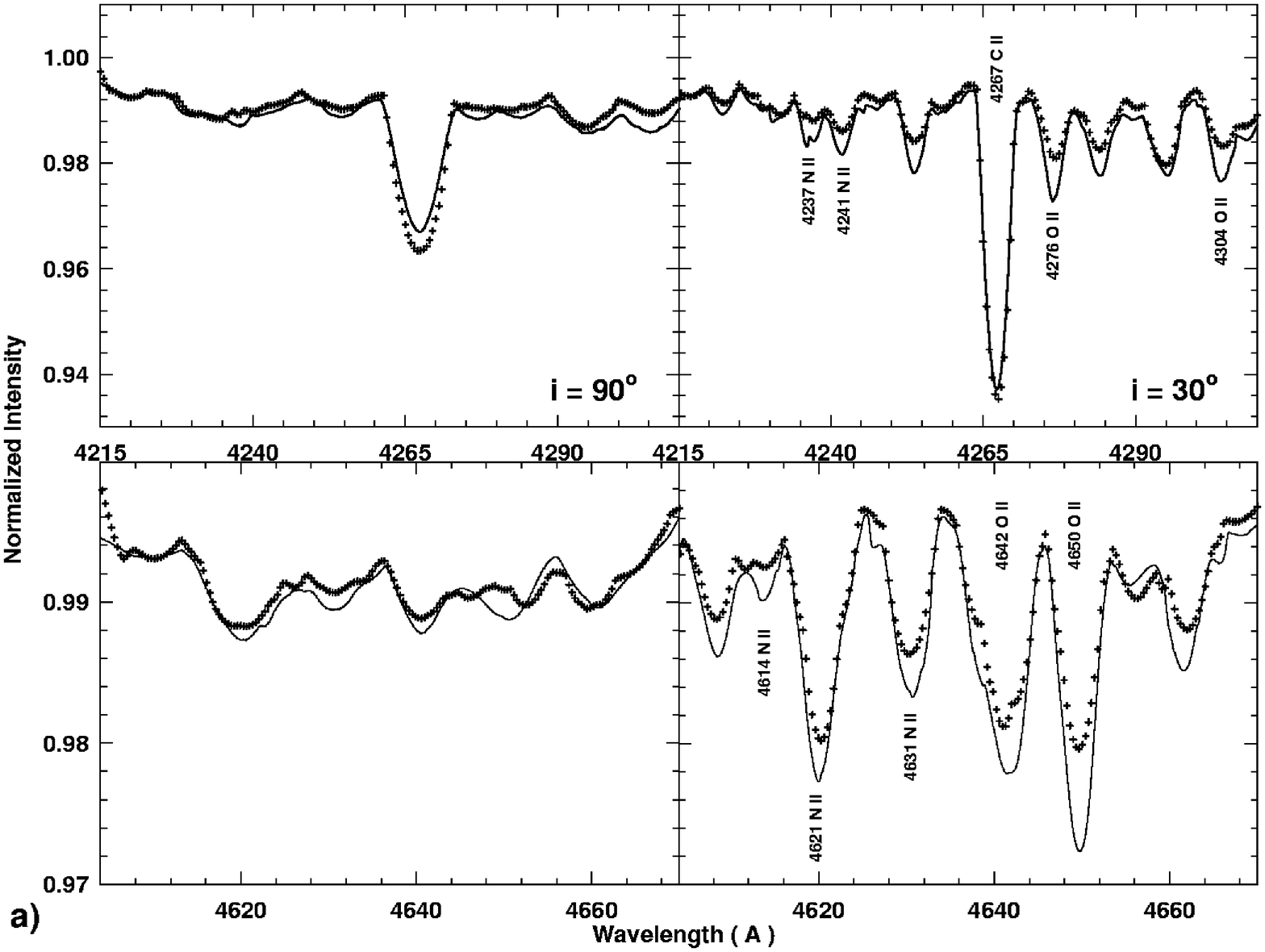}\\
\includegraphics[width=16cm,angle=0,clip=]{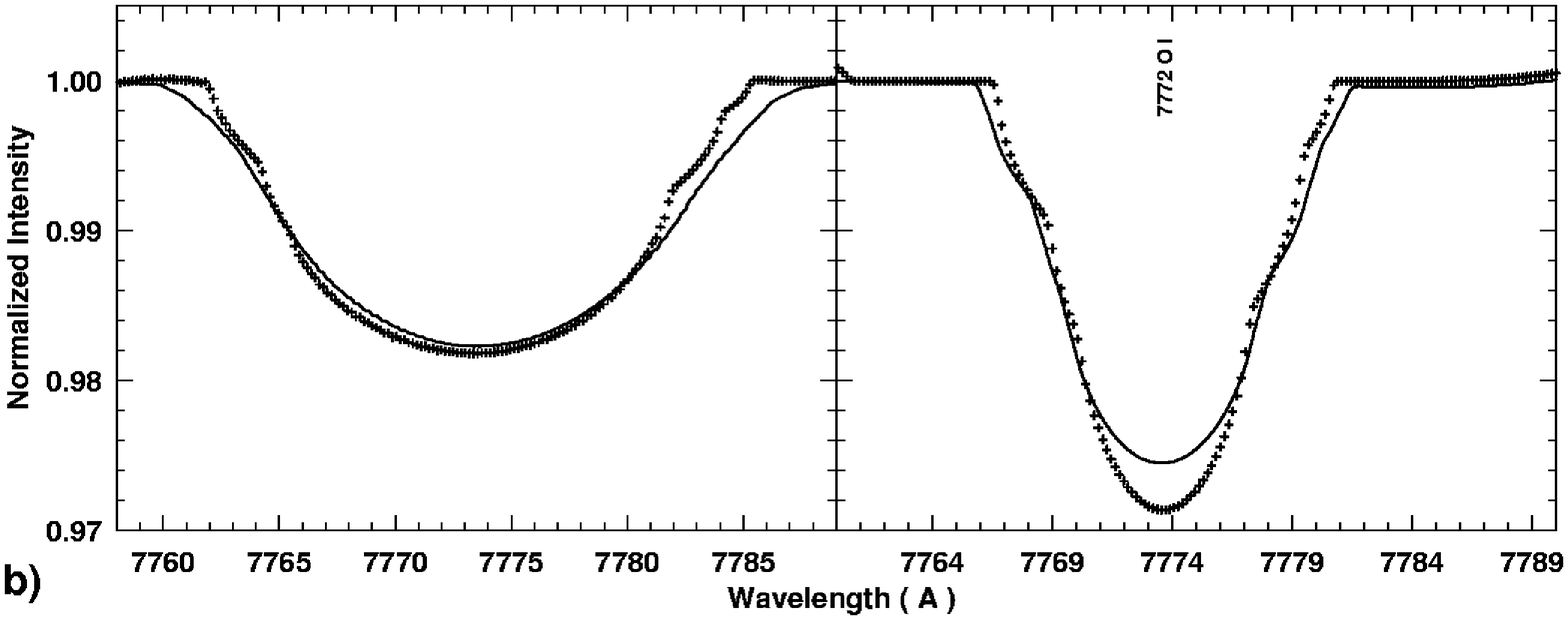}
\end{tabular}
\caption{Comparison between spectra computed using pnrc (full
line) and apparent parameters (crosses). Pnrc spectra were
computed for \top~=~20000~K, \gop~=~4.0 and \omc~=~0.99. Three
different inclination angles were considered. Apparent values for
each case are given in Table~\ref{tab:valap}.} \label{fig:r2}
\end{figure*}

 It is very important to know whether plane parallel model
atmospheres calculated for the apparent fundamental parameters
can describe accurately enough the whole observed spectrum of a
rotating star. This matters not only for the study of the stellar
chemical composition, but also for the energy distribution
representation, which enters frequently the fundamental parameter
determination. This is important, in particular, for fast rotating
early type stars, whose evolution is faster than that of less massive
stars, so that they are found in different evolutionary stages and
can be used to test models of stellar evolution.\par

\subsection{Chemical composition}

 In order to { verify} the reliability of the use of
computed stellar spectra that ignore GD, we studied wavelength
domains that contain several helium and hydrogen lines, as well
as CNO spectral lines. We compared apparent spectra, i.e. those
computed with a plane-parallel model atmosphere using the apparent
fundamental parameters given in Table~\ref{tab:valap} with pnrc
spectra (i.e. those calculated using {\sc fastrot} for
\top~=20000~K, \gop~=4) and several inclination angles
\omc~=~0.99. This comparison for hydrogen and helium lines is
shown in Fig.~\ref{fig:r1}. The same type of comparison for CNO
elements is shown in Fig.~\ref{fig:r2}. It appears then that in
general REFs can be ignored for stars with \vsini$\la$150, which
correspond either to slow rotators or pole-on fast-rotators.
However, in fast rotating stars seen at high inclination angles,
ie: large \vsini~values, discrepancies between both series of
models can be significant. Thus, the chemical composition of
objects seen at high inclinations, which are also the most
frequently found, must be studied using models adapted for
rotating stars.\par
 In panel (a) of Fig. \ref{fig:r1} are shown the helium and
hydrogen lines used to determine the apparent \teff\ and \logg,
as noted in Sect.~\ref{sec:proc}. The spectral regions where the
fit was actually performed are marked on top of spectra with
horizontal thick lines. In general, these same plane-parallel
models calculated and apparent fundamental parameters are
representing quite satisfactorily the helium and hydrogen lines in
other spectral regions (see panel (b) in Fig.~\ref{fig:r1}).\par
\begin{figure}[ht!]
\center
\includegraphics[width=8.5cm,angle=0,clip=]{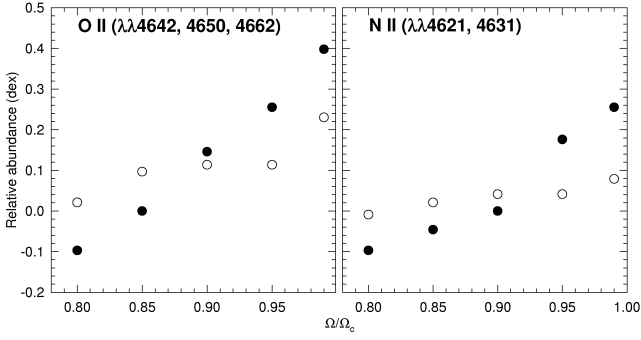}
\caption{Abundance overestimation in dex due to stellar flattening
and gravitational darkening observed for different values of
\omc~and \vsiniap~equal to 150~\kps~(open circles) and
200~\kps~(filled circles). Adopted fundamental parameters are
listed in Table~\ref{tab:abund1}. Spectral lines used in the
abundance estimate are noted in each Figure.} \label{fig:abund}
\end{figure}

 The behaviour of the CNO spectral lines as function of the
inclination angle and angular velocity is however more complex,
as shown in Fig. \ref{fig:r2}. In Fig.~\ref{fig:abund}, we
reported the overabundance of oxygen and nitrogen derived for
models corresponding to an {\it apparent} B2 IV star (\tap=19000~K
and \gap=3.5) but for different values of \omc\ and two different
\vsini~values (150~\kps~and 200~\kps). The adopted fundamental
parameters are listed in Table~\ref{tab:abund1}. This kind of
modeling shows that in early B-type stars C {\sc ii} is expected
to be underestimated when use is made of plane-parallel model
atmosphere for fast rotators seen equator-on. Nevertheless, these
models can still be used when stars are seen at low inclinations.
The abundance of nitrogen, as derived from various N {\sc ii}
lines, is overestimated by a plane-parallel modeling of fast
rotators seen at intermediate inclination angles. In fast
rotating B-type stars cooler than 25000~K, oxygen is always
overestimated by plane-parallel models when fits of O{\sc ii}
spectral lines are done, while it is expected to be somewhat
underestimated if one uses the $\lambda$7772~O{\sc i} triplet.\par

\begin{table}
\center \caption{{\it Apparent} (\tap, \gap, \vsiniap), {\it pnrc}
(\top, \gop) parameters and true \vsini\ used to study the effects
of stellar flattening on the determination of CNO abundances
(Fig.~\ref{fig:abund}).}
\begin{tabular}{@{}l@{~~}l@{~}l@{~}l@{~}l@{~~}l@{~}l@{~}l@{}}
\hline\noalign{\smallskip}
\tap & \gap & \vsiniap & \omc & \top & \gop & \vsini & i \\
 (K) &      & (\kps)  &     & (K)  &   & (\kps) & ($^{\rm o}$)\\\noalign{\smallskip}
 \hline\noalign{\smallskip}
19000. & 3.5  & 150 & 0.80 & 19433 & 3.64 & 158 & 38\\
19000. & 3.5  & 150 & 0.85 & 19709 & 3.62 & 160 & 34\\
19000. & 3.5  & 150 & 0.90 & 19662 & 3.65 & 161 & 30\\
19000. & 3.5  & 150 & 0.95 & 19537 & 3.66 & 164 & 27\\
19000. & 3.5  & 150 & 0.99 & 20480 & 3.74 & 170 & 24\\
19000. & 3.5  & 200 & 0.80 & 19529 & 3.66 & 207 & 50\\
19000. & 3.5  & 200 & 0.85 & 19680 & 3.69 & 210 & 45\\
19000. & 3.5  & 200 & 0.90 & 20070 & 3.71 & 212 & 38\\
19000. & 3.5  & 200 & 0.95 & 20424 & 3.75 & 218 & 35\\
19000. & 3.5  & 200 & 0.99 & 20818 & 3.75 & 227 &
32\\\noalign{\smallskip} \hline
\end{tabular}
\label{tab:abund1}
\end{table}

 More generally, spectral lines preferably formed at the stellar
poles (N~{\sc ii}, O~{\sc ii} ...) tend to appear stronger when
the effects of fast rotation are taken into account { (see f.e.
\ion{O}{ii} lines in panel (a) of Fig. \ref{fig:r2})}. On the
contrary, when their privileged forming region is near the
equator, they often appear weaker { (see f.e. \ion{O}{i} line in
panel (b) of Fig. \ref{fig:r2})} but the REFs are also generally
smaller. Therefore, REFs on chemical abundance determinations
strongly depend on the studied transition, on \vsini~ and on the
effective temperature of stars. The content of CNO elements in
atmospheres of early-type fast rotators, in particular Be stars,
will be discussed in a forthcoming paper.\par

\begin{table}[t!]
\caption{True and apparent fundamental parameters used in
Fig.~\ref{fig:r1} and \ref{fig:r2} at \omc=0.99 .}
\label{tab:valap}
\begin{tabular}{@{}rrrrrrr@{}}
\noalign{\smallskip}
\hline
\noalign{\smallskip}
i  & \top  & \gop & \vsinio & \tap & \gap & \vsiniap \\
\noalign{\smallskip}
\hline
\noalign{\smallskip}
0  & 20000 & 4.0 &   0 & 18675 & 3.83 & 0\\
30 & 20000 & 4.0 & 231 & 18462 & 3.75 & 203 \\
90 & 20000 & 4.0 & 463 & 16249 & 3.28 & 382 \\
\noalign{\smallskip}
\hline
\end{tabular}
\end{table}

\subsection{Energy distribution}

\begin{figure}[ht!]
\center
\includegraphics[width=7cm,angle=0,clip=]{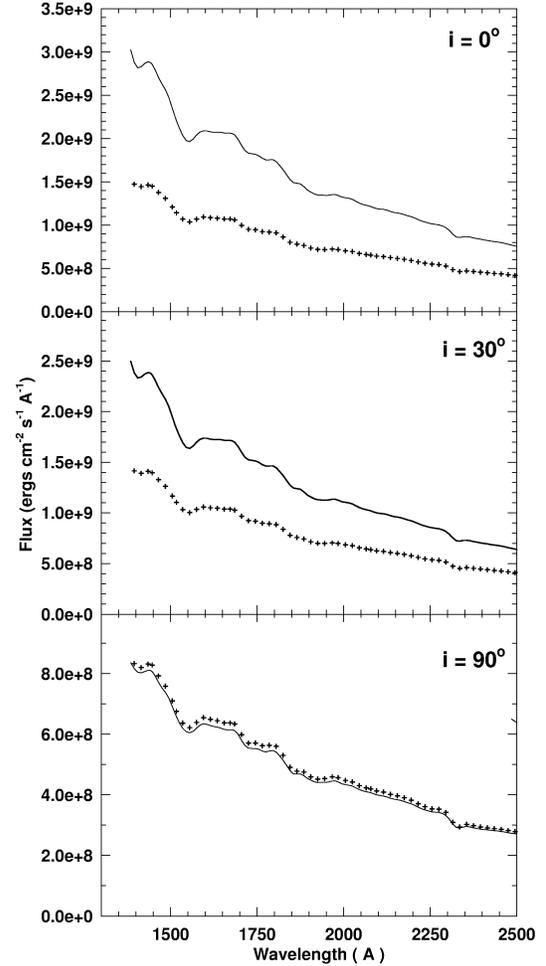}
\caption{Comparison between spectra computed using pnrc (full
line) and apparent parameters (crosses). Pnrc spectra were
computed for \top~=~20000~K, \gop~=~4.0 and \omc~=~0.99. Three
different inclination angles were considered. Apparent values for
each case are given in Table~\ref{tab:valap}.} \label{fig:td1}
\end{figure}

 Since REFs on the stellar continuum are directly proportional
to the shape and apparent size of the observed stellar disk,
magnitudes are very sensitive to the stellar flattening. The GD
displayed by the stellar hemisphere facing the oberver enhances
even more these effects. REFs affect therefore the slope and the
absolute level of the emitted fluxes. REFs on the stellar
continuum in the visible spectral range have already been widely
discussed in many previous works \citep{1963ApJ...138.1134C,
1965ApJ...142..265C,1966ApJ...146..914C,1968ApJ...151..217C,
1968ApJ...152..847C,1974ApJ...191..157C,1968ApJ...151.1057H,
1972A&A....17..161H,1968ApJ...153..465H,1971A&A....13..353H,
1977ApJS...34...41C,1991ApJS...77..541C,1970A&A.....7..120M,
1972A&A....21..279M,1985MNRAS.213..519C,1986serd.book.....Z,
1999A&A...346..586P,2004MNRAS.350..189T}. To complete the
presentation of rotational effects and because the spatial UV
spectral region was less explored in the literature
\citep{1979A&A....78..292K,1979A&A....72..318L} we show in
Fig.~\ref{fig:td1} the comparison between spatial UV energy
distribution computed with {\sc fastrot} using true and apparent
parameters for \omc~=~0.99 and different inclination angles. The
values of parameters used in each calculation are given in
Table~\ref{tab:valap}. Sometimes the spatial region of stellar
energy distributions are used to compare apparent diameters or to
infer distances \citep{2002A&A...385..986F}. According to
estimates shown in Fig.~\ref{fig:td1} differences up to 70\% are
expected in these quantities if they concern early type fast
rotators.\par

\section{Discussion}

 In the present approach of rotational effects on spectra
emitted by early-type stars we neglected second order effects due
to radiative fluxes related to the latitudinal gradient of the
local effective temperature. They might be not entirely
negligible \citep{1992A&A...256..519H}, in particular for very
fast rotators, where the equatorial gravity $\lim_{\omega\to1}
g({\rm eq})\to 0$. It is expected that their effect may reduce
somewhat the GD.\par

\subsection{Comparison with other recent calculations}

\begin{figure}
\center
\includegraphics[width=7cm,angle=0,clip=]{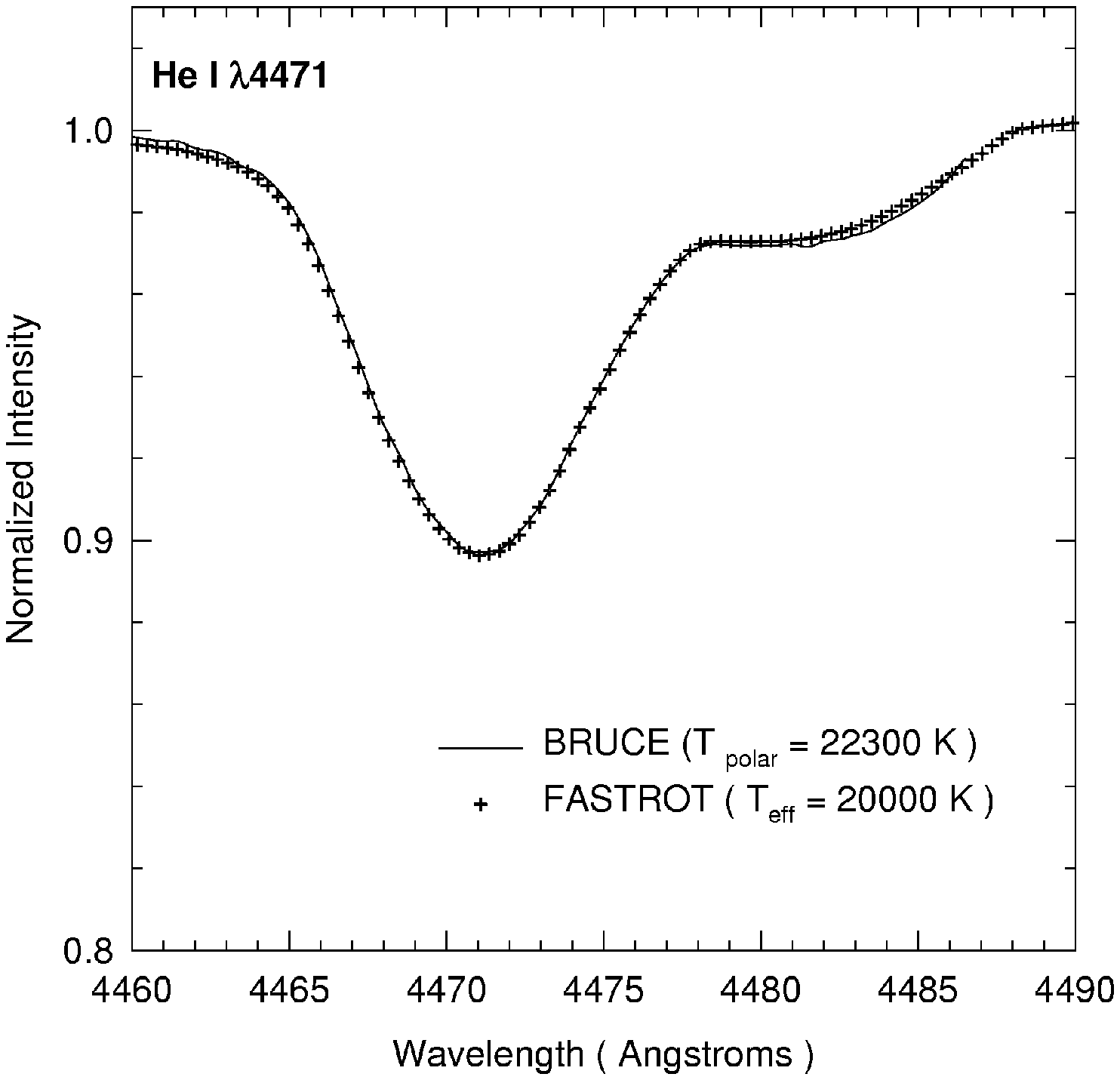}
\caption{Comparison between \ion{He}{i} $\lambda$4471 line
profiles computed with BRUCE (full line, Townsend et al. 2004)
and with FASTROT (crosses). Adopted input parameters in FASTROT
are: \top = 20000~K, \gop = 4.0, \omc\ = 0.99 and \vsinio\
=463~\kps~(ie: polar temperature = 22300~K and polar radius =
4.03~R$\odot$). Adopted input parameters in BRUCE are: polar
temperature = 22300~K, polar radius = 4.03~R$\odot$, inclination
angle = 90$^{\rm{o}}$ and equatorial velocity = 463~\kps.}
\label{fig:tbruce}
\end{figure}

{ To test our code, we compare in Fig.~\ref{fig:tbruce}} the
\ion{He}{i} $\lambda$4471 line profile computed with {\sc
fastrot} and {\sc bruce} \citep{2004MNRAS.350..189T}. The same
model atmosphere grids were used assuming near critical rotation
\omc\ = 0.99 { (\vevc\ = 0.97)} and inclination $i = 90^{\rm o}$.
Relations (\ref{eq5}) and (\ref{eq6}) were combined to find what
{ pnrc} effective temperature is implied by the T$_{\rm polar}$
value { we used as entry parameter in {\sc bruce}.}
{ This procedure leads to an almost identical gravitational
darkening and to identical line profiles as shown in Fig.
\ref{fig:tbruce}. We made several comparisons of {\sc fastrot}
and {\sc bruce} results of the type shown in Fig.
\ref{fig:tbruce}.
Although our integration algorithms is different and outnumbers
the equivalent elementary surface elements those of {\sc bruce},
quite similar results are obtained. { The only difference between
both approaches relies on the adopted fundamental parameters.} In
{\sc fastrot} a unique set of pnrc \top\ and mass $M$ is used for
the whole series of \om\ values, while {\sc bruce} does allow to
specify the temperature distribution in 2 possible ways: by
fixing T$_{\rm polar}$ or by fixing the bolometric luminosity
\citep[see eq. 3 in][]{2004MNRAS.350..189T}.

}

\subsection{Implications for the study of Be stars}

It has been pointed out by \citet{1968MNRAS.140..149S} and
\citet{2004MNRAS.350..189T} that the FWHM of the $\lambda$4471
\ion{He}{i} line can converge to a constant value at high \vsini\
when the stellar angular velocity approaches the critical
rotation, which makes that the \vsini\ values of fast rotators,
in particular those of Be stars, can be underestimated.
Nevertheless, the implication on the measurements of \vsini\ in
actual early-type stars by taking into account entirely the blend
\ion{He}{i}4470+\ion{He}{i}4471+\ion{Mg}{ii}4481, has still not
been entirely explored. Our comparisons between true and apparent
values obtained using the fitting procedure (CASE A Table
\ref{tab:fitr}), which mimic real cases of \vsini\ determination,
show that there is a progressive lowering of the \vsini\ for
increasing rotational rates. We also see that this \vsini\
underestimation does not imply necessarily a sa\-tu\-ra\-tion of
the estimated projected rotation velocity, even there is critical
rotation (see Fig. \ref{fig:vst}).\par
 It is interesting then to revisit the most probable value of
the \omc\ rate of Be stars if the REFs are taken into account.
\citet{2001A&A...378..861C} have studied a sample of 116 Be stars
and determined their \vsiniap\ parameter without taking into
account the REFs. They used, however, Stoeckley and Mihalas'
(1973) calculations of rotationally broadened lines, where the
wavelength dependent limb-darkening within the line profiles is
considered and which ensures that there is no systematic
underestimation of rotational parameters, effect that was
discussed by \citet{1995ApJ...439..860C}.
\citet{2001A&A...378..861C} have thus concluded that on average
Be stars rotate at \omc\ $\sim$ 0.8. We have studied these stars
again, and some others, and redetermined carefully their {\it
apparent} fundamental parameters by fitting their spectra with
model atmospheres without rotation. We adopted the {\it apparent}
\vsini\ determined by \citet{2001A&A...378..861C} in order to
avoid undeterminations of these quantity produced by the Fourier
method, which uses a constant limb-darkening coefficient in the
frequencies over the line profile. Using our GRIDB of synthetic
spectra, we have determined the pnrc fundamental parameters of the
program objects, in particular their \vsini. These parameters
were calculated for an average \omc\ rate, the same for all stars
and whose value was iterated. The iteration started first by
assuming \omc\ = 0.8. At each iteration step we corrected the
{\it apparent} fundamental parameters in order to obtain the pnrc
ones and redetermined, hence, the ratios \vsini/$V_{\rm c}$ of
each stars and studied its distribution. Each time we derived
from the distribution the average of \omc\ and its mode, i.e. the
most frequent or probable value of \omc. These quantities were
estimated in two ways: a) using the same method as
\citet{2001A&A...378..861C}; b) determining the average and the
mode of $V/V_{\rm c}$ by correcting the distribution of
\vsini/$V_{\rm c}$ for the $\sin i$ effect. Since the mode of a
given distribution depends on at least its first four moments,
where errors increase with the order of the moment, we iterated
the average \omc. We also started the iteration assuming all
stars rotate at \omc = 1.0. We note that each apparent effective
temperature, gravity and projected rotational velocity was
considered with its $2\sigma$ error. The interval
($X-2\sigma_X,X+2\sigma_X$) corresponding to each {\it apparent}
$X$ parameter was divided into eight parts, so that we had 9
entries for each independent parameter. This produced $9^3$
determinations of each corresponding pnrc $X$ parameter or true
\vsini. Since each distribution of the obtained pnrc parameters is
not symmetrical, we adopted its mode to represent the
corresponding most probable solution of the sought pnrc
parameter.\par
 Both iterations ended at the same values of the average \omc\ and
the mode:

\begin{equation}
\ \ \ \ \ \ \ \ \ \ \ \ \overline{\Omega/\Omega_{\rm c}} =
0.85^{+0.06}_{-0.04} \ \ \ \ (\Omega/\Omega_{\rm c})_{\sc mode} =
0.88^{+0.06}_{-0.04}
\label{modeomega}
\end{equation}

Table \ref{iterat} summarizes the characteristics of iterations.
Moments $\mu_{\rm n}$ are for $V/V_{\rm c}$ distributions. The
program stars and their fundamental parameters are given in Table
\ref{data}. In this table are given the apparent effective
temperature, surface gravity and rotational parameter with their
respective $1\sigma$ errors. {The modes of the corresponding pnrc
parameter and of the corrected \vsini\ are also listed, as well
as the critical equatorial velocity, the estimated inclination
angle and the $1\sigma$ dispersion of the respective $9^3$
determinations of each fundamental parameter.}\par
 From the rates displayed in (\ref{modeomega}) we see that critical
rotation among Be stars is reachable only by a fraction of objects
between 2 and $3\sigma$ interval beyond the $(\Omega/\Omega_{\rm
c})_{\sc mode}$. While the difference between {\it apparent} and
{\it true} \vsini\ values seems to be significant in some cases,
the increase of \omc\ is not so high, as compared to the ratio
obtained by \citet{2001A&A...378..861C} with {\it apparent}
\vsini\ parameters. This is simply due to the fact that we need
to consider also the increase from {\it apparent} to pnrc values
of the stellar fundamental parameters. These last identify
objects that are more massive and less evolved than expected from
the {\it apparent} parameters. Such increase also implies a
significant augmentation of the equatorial critical velocity
$V_{\rm c}$ which is barely overweighted by the respective {\it
true} \vsini\ value.\par

\begin{table}[ht!]
\center
\caption{Iteration of $\overline{\Omega/\Omega_{\rm c}}$ and
$(\Omega/\Omega_{\rm c})_{\sc mode}$}
\label{iterat}
\scriptsize
\begin{tabular}{@{}cccccccc@{}}
\hline
\noalign{\smallskip}
& \multicolumn{6}{c}{modes}  & \\
\cline{4-5}\\
\omc & $\overline{V/V_{\rm c}}$ & $\overline{\Omega/\Omega_{\rm c}}$ &
$V/V_{\rm c}$ & $\Omega/\Omega_{\rm c}$ & $\mu_2$ & $\mu_3$ & $\mu_4$ \\
\noalign{\smallskip}
\hline
\noalign{\smallskip}
0.800 & 0.708 & 0.842 & 0.756 & 0.879 & 0.507 & 0.369 & 0.271 \\
0.842 & 0.712 & 0.846 & 0.753 & 0.878 & 0.513 & 0.375 & 0.277 \\
0.846 & 0.713 & 0.847 & 0.746 & 0.872 & 0.515 & 0.376 & 0.277 \\
0.847 & 0.713 & 0.847 & 0.751 & 0.876 & 0.515 & 0.376 & 0.278 \\
\noalign{\smallskip}
\hline
\noalign{\smallskip}
1.000 & 0.782 & 0.898 & 0.726 & 0.854 & 0.627 & 0.513 & 0.428 \\
0.898 & 0.725 & 0.856 & 0.673 & 0.814 & 0.533 & 0.397 & 0.300 \\
0.856 & 0.715 & 0.848 & 0.733 & 0.862 & 0.518 & 0.379 & 0.281 \\
0.848 & 0.713 & 0.846 & 0.732 & 0.861 & 0.514 & 0.376 & 0.278 \\
0.846 & 0.713 & 0.847 & 0.751 & 0.876 & 0.514 & 0.375 & 0.277 \\
\noalign{\smallskip}
\hline
\end{tabular}
\end{table}

\subsubsection{Comments on the solutions obtained}

 We remind that the values of parameters displayed in Table
\ref{data} are 'modes' of the respective distributions of $9^3$
possible solutions. They can be considered as representing the
most probable configuration of fundamental parameters for the
rotating star, as they are perceived from the entry set of {\it
apparent} parameters with their uncertainties, seen through the
models calculated with {\sc fastrot} used to interpret them. They
look like a sort of independent "determinations" which also seem
sometimes to underestimate the effects due to fast rotation.
Nevertheless, if we used only the {\it apparent} entry parameters
without their uncertainties, the unique translation into pnrc
quantities that comes out would lead in many cases to $\Delta
T_{\rm eff}$, $\Delta\log g$ and $\Delta$\vsini\ differences
between pnrc and {\it apparent} values that are higher than those
issued from the values in Table \ref{data}. We calculated the
$\overline{\Omega/\Omega_{\rm c}}$ and $(\Omega/\Omega_{\rm
c})_{\rm mode}$ using these simplified estimates of pnrc
parameters and true \vsini, as well as the same iteration
procedure for $\Omega/\Omega_{\rm c}$. We obtained thus,
$\overline{\Omega/\Omega_{\rm c}} =$ 0.83, near the value put
forward in \citet{2003sf2a.conf..617Z} and $(\Omega/\Omega_{\rm
c})_{\rm mode} =$ 0.88. This small increase of the                      
$\Omega/\Omega_{\rm c}$, as compared to the value obtained by
\citet{2001A&A...378..861C}, is entirely ascribed to the higher
$V_{\rm c}$ values implied by the {\it apparent} $\to$ pnrc
transformation of fundamental parameters.\par

\subsubsection{pnrc and actual fundamental parameters}

 In this paper, we have studied the relation between the {\it
apparent} parameters and those that should represent the actual
stellar mass and its evolutionary stage. The translation obtained
into pnrc parameters was performed with the help of evolutionary
tracks calculated for non rotating stars
\citep{1992A&AS...96..269S}. However, evolutionary tracks of
rotating objects are somewhat different
\citep{2000ApJ...544.1016H, 2000A&A...361..101M}. Depending on
the case, they may produce slightly different mass estimates.
These estimates are those which we might finally consider as a
better approach to the {\it actual} stellar mass. Nevertheless,
the mentioned differences are altogether small and do not modify
whatsoever the results obtained in the preceding sections. Their
{\it ins} and {\it outs} are out of the scope of the present work
and will be widely discussed in another contribution.\par

\section{Conclusions}

 In this paper we present the calculation code {\sc fastrot} to
calculate the effects carried by fast rotation on the fundamental
parameters of early-type stars. Particular attention is payed to
represent hydrogen Balmer, He\,{\sc i} and Mg\,{\sc ii} lines in
the blue spectral region, which are currently used to infer from
spectroscopic data the fundamental parameters of early-type stars.
We also calculated a number of C, N and O lines in the visible
spectral range, which enter the CNO abundance determination.\par
 We have calculated synthetic spectra with classical plane-parallel
non-LTE model atmospheres using {\it apparent} fundamental
parameters derived from H$\gamma$, He\,{\sc i}4471 and Mg\,{\sc
ii}4481 lines and compared their predictions for other spectral
regions of rotating stars. We noted that the above {\it apparent}
fundamental reliably reproduce also the remaining Balmer and
He\,{\sc i} lines in the visual. However, discrepant fits are
expected for CNO lines. The noted discrepancies may thus lead to
systematic overestimates of the N abundances, while the C
abundances can be underestimated. Regarding the O abundance, it
depends on the lines used. They are sensitive to local formation
conditions and reflect preferences either on the polar or
equatorial formation regions. O abundances can then be under- or
overestimated, depending on the lines used, stellar apparent
configuration, if rotational effects are neglected.\par
 The fastest rotation in the main sequence are probably Be stars.
The strong gravitational darkening can lead then to sensitive
\vsini\ underestimations, which in turn may imply that the
hitherto undercritical rotation of these stars can be argued. We
payed much attention then on the determination of the fundamental
parameters of fast rotators, in particular on \vsini. We thus
obtained that although the classical \vsini\ parameter do
underestimate the stellar equatorial velocity of fast rotators,
the fact that the pnrc stellar mass is higher and the evolutionary
stage lower than expected from {\it apparent} quantities, causes
that in most cases the increase in the estimate of \vsini\ does
not overweights considerably the new $V_{\rm c}$. The study of a
sample of 130 Be stars leads thus to a most probable ratio
$V_{\rm e}/V_{\rm c} \simeq$ 0.75, or $\Omega/\Omega_{\rm c}
\simeq$ 0.88, which still imply average undercritical surface
rotation.

\begin{acknowledgements}
We are grateful to the referee Dr. I.D. Howarth and to Dr. R.
Townsend for their comments and suggestions helped to improve the
presentation of our results. YF thanks Dr. P.Lampens for hosting
him at the Royal Observatory of Belgium and acknowledges funding
from the Belgian 'Diensten van de Eerste Minister - Federale
Diensten voor We\-ten\-schap\-pelij\-ke, Technische en Culturele
Aangelegenheden' (Research project MO/33/007).
\end{acknowledgements}

\bibliographystyle{aa}
\bibliography{effets5}

\begin{thebibliography}{61}
\expandafter\ifx\csname natexlab\endcsname\relax\def\natexlab#1{#1}\fi

\bibitem[{{Barnard} {et~al.}(1969){Barnard}, {Cooper}, \&
  {Shamey}}]{1969A&A.....1...28B}
{Barnard}, A.~J., {Cooper}, J., \& {Shamey}, L.~J. 1969, \aap, 1, 28

\bibitem[{{Bodenheimer}(1971)}]{1971ApJ...167..153B}
{Bodenheimer}, P. 1971, \apj, 167, 153

\bibitem[{{Chauville} {et~al.}(2001){Chauville}, {Zorec}, {Ballereau},
  {Morrell}, {Cidale}, \& {Garcia}}]{2001A&A...378..861C}
{Chauville}, J., {Zorec}, J., {Ballereau}, D., {et~al.} 2001, \aap, 378, 861

\bibitem[{{Claret}(1998)}]{1998A&A...335..647C}
{Claret}, A. 1998, \aap, 335, 647

\bibitem[{{Claret}(2000)}]{2000A&A...359..289C}
{Claret}, A. 2000, \aap, 359, 289

\bibitem[{{Clement}(1979)}]{1979ApJ...230..230C}
{Clement}, M.~J. 1979, \apj, 230, 230

\bibitem[{{Collins}(1963)}]{1963ApJ...138.1134C}
{Collins}, G.~W. 1963, \apj, 138, 1134

\bibitem[{{Collins}(1965)}]{1965ApJ...142..265C}
{Collins}, G.~W. 1965, \apj, 142, 265

\bibitem[{{Collins}(1966)}]{1966ApJ...146..914C}
{Collins}, G.~W. 1966, \apj, 146, 914

\bibitem[{{Collins}(1968{\natexlab{a}})}]{1968ApJ...151..217C}
{Collins}, G.~W. 1968{\natexlab{a}}, \apj, 151, 217

\bibitem[{{Collins}(1968{\natexlab{b}})}]{1968ApJ...152..847C}
{Collins}, G.~W. 1968{\natexlab{b}}, \apj, 152, 847

\bibitem[{{Collins}(1974)}]{1974ApJ...191..157C}
{Collins}, G.~W. 1974, \apj, 191, 157

\bibitem[{{Collins} \& {Smith}(1985)}]{1985MNRAS.213..519C}
{Collins}, G.~W. \& {Smith}, R.~C. 1985, \mnras, 213, 519

\bibitem[{{Collins} \& {Sonneborn}(1977)}]{1977ApJS...34...41C}
{Collins}, G.~W. \& {Sonneborn}, G.~H. 1977, \apjs, 34, 41

\bibitem[{{Collins} \& {Truax}(1995)}]{1995ApJ...439..860C}
{Collins}, G.~W. \& {Truax}, R.~J. 1995, \apj, 439, 860

\bibitem[{{Collins} {et~al.}(1991){Collins}, {Truax}, \&
  {Cranmer}}]{1991ApJS...77..541C}
{Collins}, G.~W., {Truax}, R.~J., \& {Cranmer}, S.~R. 1991, \apjs, 77, 541

\bibitem[{{Cunto} {et~al.}(1993){Cunto}, {Mendoza}, {Ochsenbein}, \&
  {Zeippen}}]{1993A&A...275L...5C}
{Cunto}, W., {Mendoza}, C., {Ochsenbein}, F., \& {Zeippen}, C.~J. 1993, \aap,
  275, L5

\bibitem[{{Didelon}(1982)}]{1982A&AS...50..199D}
{Didelon}, P. 1982, \aaps, 50, 199

\bibitem[{{Endal} \& {Sofia}(1976)}]{1976ApJ...210..184E}
{Endal}, A.~S. \& {Sofia}, S. 1976, \apj, 210, 184

\bibitem[{{Endal} \& {Sofia}(1979)}]{1979ApJ...232..531E}
{Endal}, A.~S. \& {Sofia}, S. 1979, \apj, 232, 531

\bibitem[{{Fr{\' e}mat} {et~al.}(2002){Fr{\' e}mat}, {Zorec}, {Hubert},
  {Cidale}, {Rohrmann}, {D{\' e}sert}, \& {Ferlet}}]{2002A&A...385..986F}
{Fr{\' e}mat}, Y., {Zorec}, J., {Hubert}, A.-M., {et~al.} 2002, \aap, 385, 986

\bibitem[{{Gray} \& {Corbally}(1994)}]{1994AJ....107..742G}
{Gray}, R.~O. \& {Corbally}, C.~J. 1994, \aj, 107, 742

\bibitem[{{Hadrava}(1992)}]{1992A&A...256..519H}
{Hadrava}, P. 1992, \aap, 256, 519

\bibitem[{{Hardorp} \& {Scholz}(1971)}]{1971A&A....13..353H}
{Hardorp}, J. \& {Scholz}, M. 1971, \aap, 13, 353

\bibitem[{{Hardorp} \&
  {Strittmatter}(1968{\natexlab{a}})}]{1968ApJ...153..465H}
{Hardorp}, J. \& {Strittmatter}, P.~A. 1968{\natexlab{a}}, \apj, 153, 465

\bibitem[{{Hardorp} \&
  {Strittmatter}(1968{\natexlab{b}})}]{1968ApJ...151.1057H}
{Hardorp}, J. \& {Strittmatter}, P.~A. 1968{\natexlab{b}}, \apj, 151, 1057

\bibitem[{{Hardorp} \& {Strittmatter}(1972)}]{1972A&A....17..161H}
{Hardorp}, J. \& {Strittmatter}, P.~A. 1972, \aap, 17, 161

\bibitem[{{Heger} \& {Langer}(2000)}]{2000ApJ...544.1016H}
{Heger}, A. \& {Langer}, N. 2000, \apj, 544, 1016

\bibitem[{{Hubeny} \& {Lanz}(1995)}]{1995ApJ...439..875H}
{Hubeny}, I. \& {Lanz}, T. 1995, \apj, 439, 875

\bibitem[{{Kodaira} \& {Hoekstra}(1979)}]{1979A&A....78..292K}
{Kodaira}, K. \& {Hoekstra}, R. 1979, \aap, 78, 292

\bibitem[{{Kurucz}(1993)}]{cdrom13}
{Kurucz}, R.~L. 1993, Kurucz CD-ROM No.13. Cambridge, Mass.: Smithsonian
  Astrophysical Observatory.

\bibitem[{{Lanz} \& {Hubeny}(2003)}]{2003ApJS..146..417L}
{Lanz}, T. \& {Hubeny}, I. 2003, \apjs, 146, 417

\bibitem[{{Leckrone}(1971)}]{1971A&A....11..387L}
{Leckrone}, D.~S. 1971, \aap, 11, 387

\bibitem[{{Llorente de Andres} \& {Duran}(1979)}]{1979A&A....72..318L}
{Llorente de Andres}, F. \& {Duran}, C.~M. 1979, \aap, 72, 318

\bibitem[{{Lucy}(1967)}]{1967ZA.....65...89L}
{Lucy}, L.~B. 1967, Zeitschrift fur Astrophysics, 65, 89

\bibitem[{{Maeder} \& {Meynet}(1996)}]{1996A&A...313..140M}
{Maeder}, A. \& {Meynet}, G. 1996, \aap, 313, 140

\bibitem[{{Maeder} \& {Meynet}(2000)}]{2000A&A...361..159M}
{Maeder}, A. \& {Meynet}, G. 2000, \aap, 361, 159

\bibitem[{{Maeder} \& {Peytremann}(1970)}]{1970A&A.....7..120M}
{Maeder}, A. \& {Peytremann}, E. 1970, \aap, 7, 120

\bibitem[{{Maeder} \& {Peytremann}(1972)}]{1972A&A....21..279M}
{Maeder}, A. \& {Peytremann}, E. 1972, \aap, 21, 279

\bibitem[{{Meynet} \& {Maeder}(2000)}]{2000A&A...361..101M}
{Meynet}, G. \& {Maeder}, A. 2000, \aap, 361, 101

\bibitem[{{Mihalas} {et~al.}(1974){Mihalas}, {Barnard}, {Cooper}, \&
  {Smith}}]{1974ApJ...190..315M}
{Mihalas}, D., {Barnard}, A.~J., {Cooper}, J., \& {Smith}, E.~W. 1974, \apj,
  190, 315

\bibitem[{{Moss} \& {Smith}(1982)}]{1982RPPh...44..831M}
{Moss}, D. \& {Smith}, R.~C. 1982, Reports of Progress in Physics, 44, 831

\bibitem[{{P{\' e}rez Hern{\' a}ndez} {et~al.}(1999){P{\' e}rez Hern{\'
  a}ndez}, {Claret}, {Hern{\' a}ndez}, \& {Michel}}]{1999A&A...346..586P}
{P{\' e}rez Hern{\' a}ndez}, F., {Claret}, A., {Hern{\' a}ndez}, M.~M., \&
  {Michel}, E. 1999, \aap, 346, 586

\bibitem[{{Pinsonneault} {et~al.}(1991){Pinsonneault}, {Deliyannis}, \&
  {Demarque}}]{1991ApJ...367..239P}
{Pinsonneault}, M.~H., {Deliyannis}, C.~P., \& {Demarque}, P. 1991, \apj, 367,
  239

\bibitem[{{Reiners}(2003)}]{A&A...408..707R}
{Reiners}, A. 2003, \aap, 408, 707

\bibitem[{{Sackmann}(1970)}]{1970A&A.....8...76S}
{Sackmann}, I.~J. 1970, \aap, 8, 76

\bibitem[{{Schaller} {et~al.}(1992){Schaller}, {Schaerer}, {Meynet}, \&
  {Maeder}}]{1992A&AS...96..269S}
{Schaller}, G., {Schaerer}, D., {Meynet}, G., \& {Maeder}, A. 1992, \aaps, 96,
  269

\bibitem[{{Sigut}(1996)}]{1996ApJ...473..452S}
{Sigut}, T.~A.~A. 1996, \apj, 473, 452

\bibitem[{{Slettebak} {et~al.}(1975){Slettebak}, {Collins}, {Parkinson},
  {Boyce}, \& {White}}]{1975ApJS...29..137S}
{Slettebak}, A., {Collins}, G.~W., {Parkinson}, T.~D., {Boyce}, P.~B., \&
  {White}, N.~M. 1975, \apjs, 29, 137

\bibitem[{{Smith} \& {Worley}(1974)}]{1974MNRAS.167..199C}
{Smith}, C.~R. \& {Worley}, R. 1974, \mnras, 167, 199

\bibitem[{{Stoeckley}(1968)}]{1968MNRAS.140..149S}
{Stoeckley}, T.~R. 1968, \mnras, 140, 149

\bibitem[{{Tassoul}(1978)}]{1978trs..book.....T}
{Tassoul}, J. 1978, {Theory of rotating stars} (Princeton Series in
  Astrophysics, Princeton: University Press, 1978)

\bibitem[{{Townsend} {et~al.}(2004){Townsend}, {Owocki}, \&
  {Howarth}}]{2004MNRAS.350..189T}
{Townsend}, R.~H.~D., {Owocki}, S.~P., \& {Howarth}, I.~D. 2004, \mnras, 350,
  189

\bibitem[{{Tuominen}(1972)}]{1972ApL....10..175T}
{Tuominen}, J. 1972, \aplett, 10, 175

\bibitem[{{Varosi} {et~al.}(1995){Varosi}, {Lanz}, {deKoter}, {Hubeny}, , \&
  {Heap}}]{modion}
{Varosi}, F., {Lanz}, T., {deKoter}, A., {et~al.} 1995,
  ftp://idlastro.gsfc.nasa.gov/pub/contrib/varosi/modion

\bibitem[{{Vidal} {et~al.}(1973){Vidal}, {Cooper}, \&
  {Smith}}]{1973ApJS...25...37V}
{Vidal}, C.~R., {Cooper}, J., \& {Smith}, E.~W. 1973, \apjs, 25, 37

\bibitem[{{Zahn}(1992)}]{1992A&A...265..115Z}
{Zahn}, J.-P. 1992, \aap, 265, 115

\bibitem[{{Zeipel}(1924)}]{1924MNRAS..84..665Z}
{Zeipel}, H.~V. 1924, \mnras, 84, 665

\bibitem[{{Zorec}(1986)}]{1986serd.book.....Z}
{Zorec}, J. 1986, {PhD Thesis: Structure et rotation diff\'erentielle dans les
  \'etoiles B avec et sans \'emission} (Paris: Universite VII, 1986)

\bibitem[{{Zorec} {et~al.}(2003){Zorec}, {Fr{\' e}mat}, {Ballereau},
  {Chauville}, {Hubert}, {Floquet}, {Levenhagen}, \&
  {Leister}}]{2003sf2a.conf..617Z}
{Zorec}, J., {Fr{\' e}mat}, Y., {Ballereau}, D., {et~al.} 2003, in SF2A-2003:
  Semaine de l'Astrophysique Francaise, 617

\bibitem[{{Zorec} {et~al.}(1988){Zorec}, {Mochkovitch}, \&
  {Divan}}]{1988CRASM.306.1265Z}
{Zorec}, J., {Mochkovitch}, R., \& {Divan}, L. 1988, Academie des Sciences
  Paris Comptes Rendus Serie Sciences Mathematiques, 306, 1265

\end{thebibliography}

\nonstopmode
\setlongtables
\onecolumn
\begin{longtable}{cccc|ccccc}
\kill


\caption{Apparent and pnrc fundamental parameters of the studied
stars. An electronic version of the table will be made available
at the CDS.} \label{data}
\\
    & \multicolumn{3}{c|}{\it Apparent parameters} & \multicolumn{5}{c}{\it Mode of the pnrc parameters} \\
 HD & $T_{\rm eff}\pm\epsilon$ & $\log g\pm\epsilon$ & \vsini$\pm\epsilon$ & $T_{\rm eff}\pm1\sigma$ & $\log g\pm1\sigma$ &
\vsini$\pm1\sigma$ & $V_{\rm c}\pm1\sigma$ & $i^o\pm1\sigma$ \\
    144 &11956$\pm$324 & 3.383$\pm$0.055 & 125$\pm$07 & 12206$\pm$414 & 3.403$\pm$0.067 & 132$\pm$08 & 309$\pm$20 & 33.9$\pm$1.2  \\
   4180 &14438$\pm$272 & 3.284$\pm$0.041 & 195$\pm$10 & 15584$\pm$426 & 3.399$\pm$0.062 & 208$\pm$13 & 332$\pm$21 & 55.7$\pm$3.2  \\
   5394 &26431$\pm$618 & 3.800$\pm$0.070 & 432$\pm$28 & 30001$\pm$624 & 3.990$\pm$0.065 & 441$\pm$27 & 577$\pm$36 & 76.4$\pm$2.7  \\
   6811 &12478$\pm$444 & 3.092$\pm$0.068 &  85$\pm$05 & 12578$\pm$481 & 3.110$\pm$0.073 &  93$\pm$06 & 274$\pm$19 & 26.5$\pm$1.0  \\
  10516 &25556$\pm$659 & 3.899$\pm$0.078 & 440$\pm$30 & 29031$\pm$751 & 4.168$\pm$0.087 & 462$\pm$33 & 590$\pm$42 & 72.5$\pm$3.9  \\
  11606 &19484$\pm$297 & 3.653$\pm$0.037 & 280$\pm$14 & 21422$\pm$439 & 3.774$\pm$0.055 & 286$\pm$16 & 446$\pm$26 & 58.2$\pm$3.5  \\
  18552 &12715$\pm$284 & 3.456$\pm$0.048 & 286$\pm$16 & 14646$\pm$244 & 3.776$\pm$0.041 & 292$\pm$14 & 369$\pm$19 & 75.2$\pm$5.0  \\
  19243 &21243$\pm$388 & 3.270$\pm$0.043 & 160$\pm$08 & 22752$\pm$409 & 3.297$\pm$0.043 & 165$\pm$09 & 387$\pm$21 & 36.2$\pm$1.3  \\
  20336 &18684$\pm$517 & 3.865$\pm$0.072 & 328$\pm$21 & 20843$\pm$558 & 4.050$\pm$0.076 & 341$\pm$23 & 487$\pm$32 & 66.7$\pm$4.0  \\
  22192 &15767$\pm$509 & 3.465$\pm$0.074 & 275$\pm$19 & 17690$\pm$299 & 3.763$\pm$0.042 & 295$\pm$15 & 397$\pm$20 & 75.0$\pm$3.1  \\
  22780 &13299$\pm$284 & 3.381$\pm$0.046 & 285$\pm$15 & 15686$\pm$443 & 3.572$\pm$0.066 & 297$\pm$18 & 366$\pm$23 & 90.0$\pm$1.0  \\
  23016 &12230$\pm$241 & 3.922$\pm$0.047 & 235$\pm$12 & 13420$\pm$249 & 4.012$\pm$0.047 & 242$\pm$12 & 410$\pm$21 & 48.7$\pm$2.6  \\
  23302 &12754$\pm$504 & 3.368$\pm$0.078 & 170$\pm$12 & 13484$\pm$293 & 3.412$\pm$0.047 & 181$\pm$10 & 320$\pm$18 & 46.8$\pm$1.6  \\
  23480 &13691$\pm$481 & 3.629$\pm$0.076 & 240$\pm$16 & 14828$\pm$402 & 3.792$\pm$0.064 & 254$\pm$15 & 391$\pm$24 & 57.3$\pm$3.3  \\
  23552 &12711$\pm$416 & 3.764$\pm$0.070 & 220$\pm$14 & 13708$\pm$239 & 3.807$\pm$0.043 & 230$\pm$12 & 387$\pm$20 & 52.0$\pm$2.2  \\
  23630 &12258$\pm$505 & 3.047$\pm$0.076 & 140$\pm$10 & 12885$\pm$292 & 3.095$\pm$0.045 & 149$\pm$08 & 274$\pm$15 & 44.6$\pm$1.6  \\
  23862 &12106$\pm$272 & 3.937$\pm$0.052 & 286$\pm$16 & 13436$\pm$257 & 4.197$\pm$0.050 & 290$\pm$15 & 420$\pm$22 & 65.3$\pm$3.9  \\
  24534 &25191$\pm$513 & 3.618$\pm$0.056 & 293$\pm$17 & 26831$\pm$699 & 3.854$\pm$0.081 & 294$\pm$20 & 480$\pm$34 & 54.7$\pm$2.6  \\
  25940 &16158$\pm$582 & 3.572$\pm$0.084 & 197$\pm$14 & 17593$\pm$338 & 3.789$\pm$0.049 & 220$\pm$13 & 386$\pm$21 & 38.3$\pm$4.5  \\
  28497 &26724$\pm$427 & 4.200$\pm$0.046 & 335$\pm$17 & 28101$\pm$738 & 4.222$\pm$0.088 & 342$\pm$24 & 631$\pm$45 & 45.4$\pm$2.0  \\
  30076 &20488$\pm$330 & 3.772$\pm$0.041 & 213$\pm$11 & 21311$\pm$397 & 3.792$\pm$0.049 & 225$\pm$12 & 456$\pm$25 & 38.4$\pm$1.7  \\
  32343 &16131$\pm$514 & 3.798$\pm$0.077 &  95$\pm$06 & 16121$\pm$430 & 3.790$\pm$0.065 & 103$\pm$06 & 410$\pm$26 & 19.7$\pm$0.8  \\
  33328 &21137$\pm$599 & 3.447$\pm$0.075 & 318$\pm$22 & 24340$\pm$644 & 3.608$\pm$0.076 & 333$\pm$23 & 440$\pm$30 & 73.0$\pm$2.4  \\
  35411 &23652$\pm$510 & 3.711$\pm$0.059 & 170$\pm$10 & 25138$\pm$441 & 3.719$\pm$0.046 & 174$\pm$09 & 484$\pm$26 & 29.0$\pm$1.3  \\
  35439 &22134$\pm$665 & 3.920$\pm$0.087 & 263$\pm$19 & 23547$\pm$364 & 3.948$\pm$0.041 & 266$\pm$13 & 513$\pm$26 & 42.2$\pm$2.3  \\
  36576 &22618$\pm$508 & 3.804$\pm$0.062 & 265$\pm$16 & 24502$\pm$399 & 3.839$\pm$0.043 & 266$\pm$13 & 487$\pm$25 & 46.3$\pm$2.2  \\
  37202 &19310$\pm$550 & 3.732$\pm$0.075 & 310$\pm$7 & 21460$\pm$420 & 3.910$\pm$0.060 & 326$\pm$7 & 466$\pm$13 & 66.0$\pm$4.1  \\
  37657 &17951$\pm$542 & 3.659$\pm$0.076 & 198$\pm$13 & 19275$\pm$301 & 3.686$\pm$0.039 & 209$\pm$10 & 427$\pm$21 & 41.9$\pm$1.9  \\
  37795 &12963$\pm$203 & 3.517$\pm$0.036 & 180$\pm$09 & 13695$\pm$437 & 3.559$\pm$0.069 & 192$\pm$12 & 355$\pm$23 & 44.9$\pm$2.1  \\
  37967 &16543$\pm$264 & 3.850$\pm$0.041 & 210$\pm$10 & 17558$\pm$589 & 3.873$\pm$0.085 & 228$\pm$16 & 455$\pm$33 & 42.1$\pm$2.1  \\
  38010 &25826$\pm$448 & 4.089$\pm$0.050 & 370$\pm$20 & 28015$\pm$555 & 4.194$\pm$0.061 & 377$\pm$22 & 608$\pm$36 & 53.0$\pm$3.4  \\
  40978 &18227$\pm$571 & 3.576$\pm$0.078 & 200$\pm$14 & 19605$\pm$550 & 3.608$\pm$0.073 & 211$\pm$14 & 401$\pm$27 & 43.3$\pm$1.9  \\
  41335 &20902$\pm$610 & 3.886$\pm$0.081 & 358$\pm$25 & 23431$\pm$639 & 4.076$\pm$0.082 & 376$\pm$26 & 520$\pm$36 & 68.5$\pm$3.8  \\
  42545 &15764$\pm$334 & 3.849$\pm$0.053 & 285$\pm$16 & 17522$\pm$309 & 3.976$\pm$0.046 & 303$\pm$16 & 460$\pm$24 & 60.1$\pm$4.0  \\
  44458 &22914$\pm$532 & 3.600$\pm$0.063 & 242$\pm$15 & 24745$\pm$667 & 3.640$\pm$0.079 & 254$\pm$18 & 449$\pm$31 & 48.0$\pm$2.1  \\
  45314 &29160$\pm$467 & 3.949$\pm$0.044 & 285$\pm$14 & 31092$\pm$557 & 3.968$\pm$0.053 & 291$\pm$16 & 594$\pm$33 & 40.6$\pm$1.9  \\
  45542 &13711$\pm$526 & 3.627$\pm$0.082 & 217$\pm$15 & 14651$\pm$229 & 3.682$\pm$0.038 & 226$\pm$11 & 369$\pm$18 & 52.5$\pm$2.6  \\
  45725 &17810$\pm$455 & 3.895$\pm$0.066 & 330$\pm$20 & 19601$\pm$459 & 4.168$\pm$0.065 & 345$\pm$21 & 486$\pm$30 & 66.6$\pm$3.9  \\
  45910 &20659$\pm$484 & 2.739$\pm$0.052 & 240$\pm$14 & 22983$\pm$471 & 2.992$\pm$0.048 & 254$\pm$14 & 342$\pm$20 & 76.5$\pm$3.5  \\
  45995 &21198$\pm$632 & 3.769$\pm$0.083 & 255$\pm$18 & 22674$\pm$423 & 3.806$\pm$0.050 & 260$\pm$14 & 470$\pm$26 & 46.8$\pm$2.1  \\
  47054 &12336$\pm$481 & 3.538$\pm$0.078 & 219$\pm$15 & 13292$\pm$355 & 3.607$\pm$0.058 & 227$\pm$13 & 351$\pm$21 & 57.2$\pm$2.2  \\
  48917 &20341$\pm$435 & 3.395$\pm$0.053 & 205$\pm$12 & 21348$\pm$569 & 3.438$\pm$0.070 & 212$\pm$14 & 389$\pm$26 & 45.4$\pm$2.0  \\
  50013 &24627$\pm$590 & 4.017$\pm$0.072 & 243$\pm$15 & 25790$\pm$713 & 4.026$\pm$0.087 & 244$\pm$17 & 535$\pm$39 & 37.3$\pm$1.9  \\
  50083 &20908$\pm$604 & 3.678$\pm$0.078 & 170$\pm$12 & 21216$\pm$625 & 3.689$\pm$0.081 & 176$\pm$12 & 452$\pm$32 & 32.5$\pm$1.4  \\
  53974 &27808$\pm$759 & 3.650$\pm$0.085 & 100$\pm$17 & 27624$\pm$577 & 3.649$\pm$0.060 & 107$\pm$06 & 490$\pm$30 & 16.5$\pm$0.6  \\
  54309 &20859$\pm$397 & 3.587$\pm$0.048 & 195$\pm$10 & 22235$\pm$543 & 3.610$\pm$0.066 & 201$\pm$13 & 425$\pm$27 & 38.4$\pm$1.5  \\
  56014 &18544$\pm$614 & 3.287$\pm$0.081 & 280$\pm$20 & 21061$\pm$453 & 3.514$\pm$0.055 & 290$\pm$17 & 389$\pm$23 & 72.6$\pm$4.2  \\
  56139 &19537$\pm$331 & 3.615$\pm$0.042 &  85$\pm$04 & 19465$\pm$300 & 3.612$\pm$0.037 &  89$\pm$04 & 403$\pm$20 & 16.6$\pm$0.7  \\
  57219 &19865$\pm$575 & 3.874$\pm$0.078 &  80$\pm$05 & 19759$\pm$486 & 3.862$\pm$0.066 &  84$\pm$05 & 443$\pm$28 & 14.3$\pm$0.7  \\
  58050 &19961$\pm$465 & 3.934$\pm$0.063 & 130$\pm$08 & 19998$\pm$590 & 3.926$\pm$0.081 & 136$\pm$09 & 481$\pm$34 & 21.9$\pm$1.2  \\
  58343 &16531$\pm$409 & 3.619$\pm$0.059 &  43$\pm$02 & 16389$\pm$437 & 3.611$\pm$0.063 &  49$\pm$03 & 377$\pm$24 &  9.7$\pm$0.3  \\
  58715 &11772$\pm$344 & 3.811$\pm$0.062 & 230$\pm$14 & 12769$\pm$395 & 3.858$\pm$0.068 & 231$\pm$14 & 380$\pm$24 & 54.5$\pm$2.9  \\
  58978 &24445$\pm$476 & 4.153$\pm$0.057 & 370$\pm$21 & 26622$\pm$717 & 4.203$\pm$0.088 & 375$\pm$26 & 615$\pm$44 & 55.2$\pm$3.0  \\
  60606 &18030$\pm$332 & 3.662$\pm$0.045 & 273$\pm$14 & 19600$\pm$414 & 3.783$\pm$0.055 & 285$\pm$16 & 441$\pm$25 & 58.2$\pm$3.6  \\
  60848 &27685$\pm$671 & 4.051$\pm$0.077 & 247$\pm$16 & 28491$\pm$630 & 4.057$\pm$0.070 & 256$\pm$16 & 582$\pm$37 & 35.4$\pm$1.8  \\
  63462 &26558$\pm$635 & 3.598$\pm$0.070 & 440$\pm$29 & 30685$\pm$728 & 3.835$\pm$0.077 & 514$\pm$31 & 545$\pm$37 & 90.0$\pm$1.0  \\
  65875 &20205$\pm$532 & 3.845$\pm$0.071 & 153$\pm$10 & 20358$\pm$590 & 3.844$\pm$0.079 & 163$\pm$11 & 447$\pm$31 & 27.8$\pm$1.2  \\
  68980 &25126$\pm$642 & 3.951$\pm$0.077 & 145$\pm$09 & 25172$\pm$492 & 3.945$\pm$0.056 & 152$\pm$08 & 534$\pm$31 & 22.2$\pm$1.2  \\
  75311 &16943$\pm$429 & 3.846$\pm$0.064 & 268$\pm$16 & 18513$\pm$527 & 4.089$\pm$0.077 & 283$\pm$19 & 467$\pm$31 & 54.2$\pm$4.0  \\
  77320 &19555$\pm$488 & 3.871$\pm$0.067 & 345$\pm$22 & 21539$\pm$637 & 4.015$\pm$0.085 & 349$\pm$25 & 501$\pm$36 & 68.7$\pm$3.6  \\
  83953 &15187$\pm$329 & 3.661$\pm$0.051 & 260$\pm$14 & 16616$\pm$273 & 3.788$\pm$0.041 & 276$\pm$14 & 407$\pm$21 & 61.0$\pm$4.0  \\
  86612 &16374$\pm$577 & 3.580$\pm$0.083 & 185$\pm$13 & 17392$\pm$381 & 3.790$\pm$0.055 & 206$\pm$12 & 381$\pm$22 & 43.3$\pm$4.2  \\
  88661 &21527$\pm$384 & 3.989$\pm$0.049 & 237$\pm$12 & 22334$\pm$450 & 4.003$\pm$0.057 & 243$\pm$14 & 511$\pm$29 & 39.0$\pm$2.1  \\
  89080 &11720$\pm$431 & 3.339$\pm$0.070 & 240$\pm$16 & 13275$\pm$251 & 3.581$\pm$0.043 & 245$\pm$13 & 320$\pm$17 & 70.8$\pm$3.0  \\
  91120 &11120$\pm$428 & 3.881$\pm$0.076 & 295$\pm$21 & 12592$\pm$244 & 4.171$\pm$0.050 & 310$\pm$16 & 407$\pm$21 & 70.7$\pm$2.7  \\
  91465 &17389$\pm$415 & 3.518$\pm$0.057 & 266$\pm$16 & 19338$\pm$582 & 3.704$\pm$0.079 & 285$\pm$20 & 401$\pm$28 & 67.3$\pm$3.7  \\
 105435 &22360$\pm$589 & 3.916$\pm$0.075 & 260$\pm$17 & 23384$\pm$446 & 3.942$\pm$0.053 & 263$\pm$14 & 527$\pm$29 & 41.6$\pm$2.3  \\
 105521 &16371$\pm$451 & 3.097$\pm$0.060 & 120$\pm$07 & 17490$\pm$409 & 3.124$\pm$0.052 & 128$\pm$07 & 320$\pm$19 & 33.4$\pm$1.1  \\
 109387 &13982$\pm$392 & 3.479$\pm$0.061 & 200$\pm$12 & 15383$\pm$357 & 3.582$\pm$0.054 & 209$\pm$12 & 362$\pm$21 & 50.0$\pm$2.6  \\
 110432 &20324$\pm$344 & 3.638$\pm$0.042 & 400$\pm$21 & 24070$\pm$603 & 3.950$\pm$0.074 & 419$\pm$28 & 485$\pm$32 & 85.4$\pm$0.3  \\
 112078 &16484$\pm$409 & 4.028$\pm$0.063 & 302$\pm$18 & 18398$\pm$331 & 4.281$\pm$0.051 & 327$\pm$17 & 506$\pm$27 & 56.3$\pm$3.6  \\
 112091 &20349$\pm$541 & 3.934$\pm$0.073 & 210$\pm$14 & 21649$\pm$638 & 3.944$\pm$0.084 & 221$\pm$15 & 492$\pm$35 & 36.0$\pm$1.9  \\
 120324 &22554$\pm$661 & 3.906$\pm$0.085 & 159$\pm$11 & 22698$\pm$533 & 3.904$\pm$0.067 & 166$\pm$10 & 508$\pm$32 & 25.7$\pm$1.3  \\
 120991 &22214$\pm$631 & 3.685$\pm$0.079 &  70$\pm$04 & 22063$\pm$414 & 3.679$\pm$0.049 &  75$\pm$04 & 429$\pm$23 & 13.0$\pm$0.5  \\
 124367 &17508$\pm$429 & 3.757$\pm$0.061 & 295$\pm$18 & 19720$\pm$310 & 3.914$\pm$0.041 & 318$\pm$16 & 446$\pm$22 & 62.9$\pm$5.1  \\
 127972 &19798$\pm$525 & 3.846$\pm$0.071 & 310$\pm$20 & 21761$\pm$565 & 3.969$\pm$0.074 & 326$\pm$21 & 500$\pm$33 & 57.8$\pm$3.8  \\
 131492 &18353$\pm$576 & 3.535$\pm$0.078 & 185$\pm$13 & 19660$\pm$392 & 3.564$\pm$0.050 & 195$\pm$11 & 395$\pm$22 & 41.0$\pm$1.9  \\
 135734 &12705$\pm$240 & 3.737$\pm$0.044 & 278$\pm$14 & 14483$\pm$383 & 4.005$\pm$0.064 & 282$\pm$17 & 391$\pm$24 & 68.7$\pm$4.0  \\
 137387 &22920$\pm$456 & 4.162$\pm$0.058 & 250$\pm$14 & 23719$\pm$550 & 4.167$\pm$0.069 & 251$\pm$16 & 573$\pm$36 & 34.6$\pm$1.8  \\
 138749 &14457$\pm$401 & 3.745$\pm$0.064 & 340$\pm$21 & 16808$\pm$566 & 3.932$\pm$0.084 & 345$\pm$24 & 441$\pm$31 & 79.7$\pm$3.8  \\
 142184 &21246$\pm$397 & 4.200$\pm$0.053 & 340$\pm$18 & 22649$\pm$640 & 4.238$\pm$0.085 & 342$\pm$24 & 574$\pm$41 & 51.5$\pm$2.4  \\
 142926 &12076$\pm$403 & 3.917$\pm$0.071 & 336$\pm$22 & 13866$\pm$339 & 4.188$\pm$0.061 & 338$\pm$20 & 424$\pm$25 & 73.8$\pm$2.0  \\
 142983 &17645$\pm$554 & 3.845$\pm$0.080 & 390$\pm$27 & 20104$\pm$389 & 4.041$\pm$0.053 & 407$\pm$22 & 501$\pm$28 & 67.4$\pm$3.7  \\
 148184 &28783$\pm$727 & 3.913$\pm$0.081 & 144$\pm$10 & 28762$\pm$611 & 3.908$\pm$0.065 & 151$\pm$09 & 580$\pm$36 & 20.0$\pm$0.9  \\
 149438 &28959$\pm$526 & 3.739$\pm$0.051 &  15$\pm$03 & 28590$\pm$610 & 3.732$\pm$0.063 &  15$\pm$03 & 536$\pm$33 &  2.3$\pm$0.1  \\
 149757 &26378$\pm$729 & 3.795$\pm$0.086 & 340$\pm$25 & 28797$\pm$638 & 4.047$\pm$0.070 & 352$\pm$22 & 534$\pm$34 & 60.7$\pm$3.3  \\
 157042 &21787$\pm$533 & 4.062$\pm$0.070 & 340$\pm$22 & 23711$\pm$639 & 4.310$\pm$0.084 & 348$\pm$24 & 572$\pm$40 & 53.1$\pm$3.7  \\
 158427 &18044$\pm$310 & 3.991$\pm$0.046 & 290$\pm$15 & 19586$\pm$327 & 4.232$\pm$0.047 & 305$\pm$15 & 477$\pm$24 & 51.3$\pm$3.2  \\
 162732 &13523$\pm$243 & 3.754$\pm$0.043 & 310$\pm$16 & 15448$\pm$560 & 3.975$\pm$0.087 & 321$\pm$23 & 431$\pm$31 & 72.9$\pm$2.6  \\
 164284 &21609$\pm$523 & 3.943$\pm$0.068 & 280$\pm$17 & 22822$\pm$670 & 3.978$\pm$0.087 & 287$\pm$21 & 513$\pm$37 & 48.0$\pm$2.4  \\
 167128 &17396$\pm$608 & 3.645$\pm$0.086 &  48$\pm$03 & 17254$\pm$575 & 3.637$\pm$0.082 &  55$\pm$03 & 389$\pm$28 & 10.6$\pm$0.3  \\
 173948 &22075$\pm$472 & 3.599$\pm$0.056 & 140$\pm$08 & 22212$\pm$517 & 3.606$\pm$0.062 & 150$\pm$09 & 430$\pm$27 & 27.3$\pm$1.1  \\
 174237 &17683$\pm$556 & 3.764$\pm$0.079 & 163$\pm$11 & 18458$\pm$441 & 3.773$\pm$0.061 & 173$\pm$10 & 425$\pm$26 & 33.2$\pm$1.5  \\
 175869 &12241$\pm$496 & 3.254$\pm$0.077 & 175$\pm$12 & 13442$\pm$441 & 3.474$\pm$0.069 & 182$\pm$12 & 311$\pm$20 & 52.0$\pm$2.1  \\
 178175 &18939$\pm$286 & 3.489$\pm$0.035 & 105$\pm$05 & 18976$\pm$561 & 3.493$\pm$0.074 & 111$\pm$07 & 396$\pm$27 & 22.4$\pm$1.1  \\
 183656 &12626$\pm$524 & 3.295$\pm$0.081 & 270$\pm$20 & 14642$\pm$360 & 3.551$\pm$0.056 & 284$\pm$16 & 354$\pm$21 & 76.7$\pm$4.9  \\
 183914 &13218$\pm$445 & 3.830$\pm$0.074 & 220$\pm$14 & 14422$\pm$278 & 3.867$\pm$0.048 & 228$\pm$12 & 387$\pm$21 & 48.7$\pm$2.3  \\
 184915 &27076$\pm$540 & 3.492$\pm$0.054 & 229$\pm$13 & 28888$\pm$606 & 3.528$\pm$0.060 & 244$\pm$15 & 479$\pm$29 & 41.1$\pm$2.5  \\
 185037 &12058$\pm$212 & 3.768$\pm$0.042 & 280$\pm$14 & 13586$\pm$523 & 4.002$\pm$0.086 & 287$\pm$20 & 410$\pm$29 & 68.4$\pm$3.6  \\
 187811 &18086$\pm$583 & 3.810$\pm$0.083 & 245$\pm$17 & 19682$\pm$336 & 3.849$\pm$0.044 & 258$\pm$13 & 434$\pm$22 & 48.9$\pm$2.4  \\
 189687 &18106$\pm$379 & 3.455$\pm$0.050 & 200$\pm$11 & 19336$\pm$319 & 3.498$\pm$0.039 & 212$\pm$11 & 395$\pm$20 & 46.6$\pm$2.3  \\
 191610 &18353$\pm$516 & 3.718$\pm$0.072 & 300$\pm$20 & 20671$\pm$587 & 3.983$\pm$0.079 & 318$\pm$22 & 442$\pm$30 & 63.7$\pm$3.9  \\
 192044 &12570$\pm$490 & 3.366$\pm$0.077 & 245$\pm$17 & 14452$\pm$221 & 3.573$\pm$0.036 & 248$\pm$12 & 332$\pm$16 & 71.1$\pm$3.1  \\
 193911 &12467$\pm$424 & 3.090$\pm$0.065 & 160$\pm$10 & 13434$\pm$352 & 3.194$\pm$0.053 & 168$\pm$10 & 292$\pm$17 & 49.1$\pm$2.0  \\
 194335 &22120$\pm$373 & 4.071$\pm$0.047 & 360$\pm$19 & 24357$\pm$458 & 4.185$\pm$0.055 & 363$\pm$20 & 581$\pm$32 & 57.6$\pm$3.9  \\
 198183 &13925$\pm$332 & 3.167$\pm$0.049 & 125$\pm$07 & 14233$\pm$404 & 3.195$\pm$0.059 & 133$\pm$08 & 300$\pm$19 & 35.6$\pm$1.5  \\
 200120 &21750$\pm$655 & 3.784$\pm$0.085 & 379$\pm$27 & 24808$\pm$520 & 3.966$\pm$0.061 & 387$\pm$23 & 530$\pm$32 & 73.2$\pm$3.2  \\
 201733 &16350$\pm$504 & 3.943$\pm$0.077 & 340$\pm$23 & 18410$\pm$375 & 4.176$\pm$0.056 & 357$\pm$20 & 481$\pm$27 & 68.4$\pm$4.8  \\
 203374 &24645$\pm$660 & 3.458$\pm$0.076 & 333$\pm$23 & 27662$\pm$651 & 3.663$\pm$0.070 & 342$\pm$22 & 477$\pm$31 & 70.7$\pm$3.7  \\
 203467 &17087$\pm$521 & 3.377$\pm$0.072 & 153$\pm$10 & 18249$\pm$575 & 3.404$\pm$0.077 & 165$\pm$11 & 356$\pm$25 & 38.2$\pm$1.4  \\
 205637 &17801$\pm$470 & 3.442$\pm$0.063 & 225$\pm$14 & 19348$\pm$351 & 3.673$\pm$0.045 & 238$\pm$12 & 395$\pm$21 & 54.7$\pm$2.7  \\
 206773 &29154$\pm$746 & 3.885$\pm$0.083 & 390$\pm$28 & 31800$\pm$547 & 4.143$\pm$0.052 & 399$\pm$21 & 617$\pm$33 & 62.1$\pm$3.6  \\
 208057 &17974$\pm$396 & 3.675$\pm$0.055 & 100$\pm$05 & 17977$\pm$510 & 3.672$\pm$0.071 & 110$\pm$07 & 391$\pm$26 & 20.6$\pm$0.9  \\
 208682 &22114$\pm$652 & 3.977$\pm$0.086 & 307$\pm$22 & 23814$\pm$617 & 4.019$\pm$0.077 & 314$\pm$21 & 525$\pm$36 & 53.3$\pm$2.9  \\
 209014 &12016$\pm$408 & 3.155$\pm$0.064 & 320$\pm$21 & 14694$\pm$509 & 3.571$\pm$0.077 & 357$\pm$21 & 343$\pm$24 & 90.0$\pm$1.0  \\
 209409 &12942$\pm$402 & 3.701$\pm$0.067 & 280$\pm$18 & 14562$\pm$514 & 3.978$\pm$0.082 & 282$\pm$20 & 391$\pm$27 & 69.9$\pm$3.4  \\
 209522 &16581$\pm$408 & 3.704$\pm$0.060 & 275$\pm$16 & 18507$\pm$279 & 3.886$\pm$0.039 & 299$\pm$14 & 430$\pm$21 & 66.6$\pm$4.3  \\
 210129 &13515$\pm$240 & 3.421$\pm$0.040 & 130$\pm$06 & 14127$\pm$444 & 3.439$\pm$0.068 & 140$\pm$09 & 326$\pm$21 & 33.8$\pm$1.4  \\
 212076 &19270$\pm$326 & 3.726$\pm$0.043 &  98$\pm$05 & 19236$\pm$419 & 3.721$\pm$0.056 & 103$\pm$06 & 439$\pm$26 & 18.6$\pm$0.9  \\
 212571 &26061$\pm$736 & 3.915$\pm$0.088 & 230$\pm$17 & 27574$\pm$645 & 3.926$\pm$0.072 & 233$\pm$15 & 558$\pm$37 & 33.6$\pm$1.7  \\
 214168 &25827$\pm$434 & 4.200$\pm$0.049 & 300$\pm$16 & 27647$\pm$688 & 4.211$\pm$0.081 & 305$\pm$21 & 619$\pm$43 & 40.0$\pm$1.7  \\
 214748 &11966$\pm$356 & 3.375$\pm$0.059 & 200$\pm$12 & 13294$\pm$437 & 3.606$\pm$0.070 & 205$\pm$13 & 317$\pm$21 & 56.9$\pm$2.6  \\
 217050 &17893$\pm$509 & 3.571$\pm$0.070 & 340$\pm$22 & 20723$\pm$612 & 3.871$\pm$0.082 & 355$\pm$25 & 443$\pm$31 & 78.5$\pm$2.1  \\
 217543 &17237$\pm$584 & 3.685$\pm$0.083 & 325$\pm$24 & 19569$\pm$366 & 3.933$\pm$0.050 & 335$\pm$18 & 440$\pm$24 & 74.9$\pm$3.9  \\
 217675 &14052$\pm$439 & 3.229$\pm$0.065 & 274$\pm$19 & 16741$\pm$469 & 3.460$\pm$0.066 & 289$\pm$17 & 343$\pm$22 & 90.0$\pm$1.0  \\
 217891 &14359$\pm$295 & 3.672$\pm$0.048 &  95$\pm$05 & 14376$\pm$449 & 3.668$\pm$0.070 & 100$\pm$06 & 367$\pm$24 & 20.7$\pm$1.0  \\
 224544 &14799$\pm$279 & 3.364$\pm$0.042 & 260$\pm$14 & 16557$\pm$460 & 3.560$\pm$0.066 & 270$\pm$17 & 373$\pm$24 & 72.8$\pm$2.6  \\
 224559 &15914$\pm$329 & 3.597$\pm$0.049 & 300$\pm$16 & 18338$\pm$412 & 3.809$\pm$0.058 & 312$\pm$18 & 427$\pm$25 & 75.8$\pm$3.7  \\
 224686 &11286$\pm$218 & 3.671$\pm$0.044 & 290$\pm$16 & 12800$\pm$281 & 3.948$\pm$0.052 & 300$\pm$16 & 380$\pm$21 & 76.1$\pm$2.7  \\
\noalign{\smallskip}
\end{longtable}

\end{document}